\documentclass[aps,prd,showpacs,nofootinbib,floats,floatfix,preprintnumbers,groupedaddress,twocolumn]{revtex4}

\usepackage{bm}
\usepackage{latexsym}
\usepackage{dcolumn}
\usepackage{amsmath,amsfonts,amssymb}
\usepackage{graphicx,epsfig}
\usepackage{amsmath}
\usepackage{fancyhdr}
\usepackage{hyperref}
\usepackage{graphicx,epstopdf}
\usepackage{upgreek}
\usepackage[font=small,labelfont=bf]{caption} 
\usepackage{ulem} 
\hypersetup{
	colorlinks   = true, 
	urlcolor     = blue, 
	linkcolor    = blue, 
	citecolor   = red 
}

\def\s{\sigma} 
\def\o{\omega} 
\def\O{\Omega}

\def\no{\nonumber}
\def\a{\alpha}
\def\b{\beta}
\def\g{\gamma}
\def\d{\delta}
\def\s{\sigma}

\def\p{\partial}

\def\na{\nabla}

\def\lie{\pounds_{l}}
\def\t{\tilde}
\def\tlie{\pounds_{\t{l}}}
\def\l{\lambda}
\def\th{\theta}
\def\k{\kappa}
\def\ie{{\it i.e.}~}
\def\M{$\mathcal{M}$~}
\def\H{$\mathcal{H}$~}
\def\S{\mathcal{S}}

\begin{document}
\title{Fluid-gravity correspondence in the scalar-tensor theory of gravity: (in)equivalence of Einstein and Jordan frames}
\author{Krishnakanta Bhattacharya$^{a,b}$\footnote{\color{blue} krishnakanta@iitg.ernet.in}, Bibhas Ranjan Majhi$^a$ \footnote {\color{blue} bibhas.majhi@iitg.ernet.in} and Douglas Singleton$^b $\footnote{\color{blue} dougs@mail.fresnostate.edu}}
\affiliation{$^a$Department of Physics, Indian Institute of Technology Guwahati, Guwahati 781039, Assam, India~.
\\
$^b$California State University Fresno, Fresno, CA 93740~.}

\date{\today}

\begin{abstract}
The duality of gravitational dynamics (projected on a null hypersurface) and of fluid dynamics is investigated for the scalar tensor (ST) theory of gravity. The description of ST gravity, in both Einstein and Jordan frames, is analyzed from fluid-gravity viewpoint. In the Einstein frame the dynamical equation for the metric leads to the Damour-Navier-Stokes (DNS) equation with an external forcing term, coming from the scalar field in ST gravity. In the Jordan frame the situation is more subtle. We observe that finding the DNS equation in this frame can lead to two pictures. In one picture, the usual DNS equation is modified by a Coriolis-like force term, which originates completely from the presence of a non-minimally coupled scalar field ($\phi$) on the gravity side. Moreover, the identified fluid variables are no longer conformally equivalent with those in the Einstein frame. However, this picture is consistent with the saturation of Kovtun-Son-Starinets (KSS) bound. In the other picture, we find the standard DNS equation (i.e. without the Coriolis-like force), with the fluid variables conformally equivalent with those in Einstein frame. But, the second picture, may not agree with the KSS bound for some values of $\phi$. We conclude by rewriting the Raychaudhuri equation and the tidal force equation in terms of the relevant parameters to demonstrate how the expansion scalar and the shear-tensor evolve in the spacetime. Although, the area law of entropy is broken in ST gravity, we show that the rewritten form of Raychaudhuri's equation correctly results in the generalized second law of black hole thermodynamics.	
\end{abstract}


\maketitle

\section{Introduction}
Despite the enormous success of Einstein's theory of general relativity (GR), there are several indications \cite{Riess:2001gk, Riess:2004nr, KNOP, Perlmutter:1998np, Tonry:2003zg, Barris:2003dq, Perlmutter:1997zf, Riess:1998cb, RIESS4} implying that Einstein's GR might not be a complete theory of gravity. There are several motivations, emerging from both theory \cite{Starobinsky:1979ty, Guth:1980zm, Linde:1981mu, Damour:2002mi} and from experiment \cite{Perlmutter:1998np, Riess:2001gk, Tonry:2003zg}, that motivate the study of alternative theories of gravity. In the literature, one can find several mature alternative theories of gravity with various motivations to study each of these theories (see the review \cite{Clifton:2011jh,Nojiri:2017ncd}). The scalar-tensor theory of gravity is one of the most popular among various alternative theories of gravity. ST theory not only combines two types of fields together which mediate gravity, but also the theory is built on the strong foundation of Einstein's GR. As a result, this theory has been the subject of intense research in recent times \cite{Callan:1985ia, EspositoFarese:2003ze, Elizalde:2004mq, Saridakis:2016ahq, Crisostomi:2016czh, Langlois:2017dyl}. Recently, we have resolved the longstanding issue related to the thermodynamic description of scalar-tensor gravity. We have obtained the thermodynamic laws in the Jordan and Einstein frames by properly defining the thermodynamic parameters. This showed the equivalence of these thermodynamic parameters in the two frames \cite{Bhattacharya:2017pqc, Bhattacharya:2018xlq}.

Apart from the connection of GR with thermodynamics, there is another intriguing connection which implies that gravity could be an emergent phenomenon rather than a fundamental force. This connection was obtained first by Damour who showed that Einstein's equation, when projected onto a null hypersurface, has the structure of the Navier-Stokes (NS) equation of fluid dynamics \cite{DAMOUR}. This projection of Einstein's equation is known as the  Damour-Navier-Stokes (DNS) equation. The DNS equation contains an extra non-linear term that does not have any fluid-dynamic interpretation. This makes the DNS equation distinct from the usual NS equation. Later, this fluid dynamic connection was again established by Price and Throne using the membrane paradigm \cite{Price:1986yy}. They referred to the equation they obtained as the Hajicek equation instead of the NS equation due to the presence of the extra non-linear term. This extra non-linear term can be removed if one replaces the convective derivative by Lie-derivative \cite{Gourgoulhon:2005ng}, or by choosing the local inertial frame as the observer's frame \cite{Padmanabhan:2010rp, Kolekar:2011gw}. After this, several works investigated various aspects of this fluid-gravity connection \cite{Parikh:1997ma, Bredberg:2011jq, Bhattacharyya:2008kq, Huang:2011he, Chirco:2011ex, Bai:2012ci, Bredberg:2010ky, Cai:2012mg, Zou:2013ix, Hu:2013lua, Cai:2011xv, Huang:2011kj, Anninos:2011zn, Ling:2013kua, Eling:2012ni, Berkeley:2012kz, Lysov:2017cmc, Wu:2013aov, Compere:2011dx, De:2019wok, De:2018zxo, Chatterjee:2010gp}. 

Most of the work mentioned above on the fluid-gravity connection are based on Einstein's GR. A question naturally arises: ``is the fluid-gravity analogy a characteristic of Einstein's gravity only, or can a similar connection be found in other alternative gravity theories?''. It is also interesting to examine how the expressions of fluid parameters (such as pressure, momentum, external force {\it etc.}) and coefficients (such as the bulk viscosity coefficient, shear viscosity coefficient {\it etc.}) get modified in these alternative gravity theory ({\it e.g.} scalar-tensor theory or the higher curvature theory). Moreover, in scalar-tensor theory, there has always been a major debate concerning the two conformally related frames in which the theory is described (the Jordan frame and the Einstein frame). The debate is whether this mathematical conformal equivalence results in the physical equivalence of the two frames (see the review \cite{Faraoni:1998qx} and also a recent work \cite{Quiros:2018ryt}). Regarding this, the recent opinion is that the Jordan and Einstein frames are physically equivalent in the classical limit. However, these two frames might not be physically equivalent in the quantum regime \cite{Kamenshchik:2014waa, Banerjee:2016lco,Ruf:2017xon} (for a very recent work in this particular direction, see \cite{Frion:2018oij}). This frame dependence issue has also been investigated in the perspective of cosmology \cite{Karam:2017zno,Bahamonde:2017kbs,Karam:2018squ}. Thus, if the DNS equation can be established in the two frames and the fluid parameters and the coefficients can be identified, it will be interesting to check whether the parameters and the coefficients are equivalent in the two frames or not.

 To answer these questions, in this paper we investigate the fluid-gravity correspondence in scalar-tensor gravity. Since, in the Jordan frame, the scalar field $\phi$ is non-minimally coupled with the Ricci-scalar $R$, we show that the usual way of obtaining the DNS equation does not work in the Jordan frame (we discuss this in more detail in our analysis below). Instead, one can obtain the gravitational fluid dynamic equation in the Jordan frame via two different routes. In one method, we obtain the gravitational DNS equation with the Coriolis-like force term. In this case, the fluid variables are not equivalent to those of the Einstein frame. However, this picture apparently agrees with some recent works in the context of anti-de Sitter/conformal field theory (AdS/CFT) correspondence (such as \cite{Brustein:2008cg}). In this case, the shear viscosity coefficient $\eta$ is obtained as $\eta=\phi/16\pi G$. As a result, the ratio of $\eta$ to entropy density ($s$) saturates the Kovtun-Son-Starinets (KSS) bound \cite{Kovtun:2004de}. We call this picture ``case 1". However, we show that there is another possible way to obtain the DNS equation, where, the fluid variables are conformally equivalent. We call this picture, ``case 2". This latter method of obtaining the DNS equation agrees with the equivalence of the two frames at the classical level and is more consistent from the thermodynamic viewpoint. In the Einstein frame, we show, that obtaining the DNS equation is similar to the usual GR case. However, one difference is that one can obtain the external force term in the DNS equation in the Einstein frame (also in the Jordan frame) even in the absence of external matter fields, unlike the case in standard GR.  After obtaining the DNS equation in the two frames, we investigate the evolution of the expansion scalar (which is given by the Raychaudhuri equation) and the shear tensor (which is given by the tidal force equation). We rewrite the Raychaudhuri equation and the tidal force equation in terms of the expansion scalar and the shear tensor. We show that the Raychaudhuri equation in the Jordan frame, when it is written in terms of the parameters defined for the equivalent fluid-dynamic description (case 2), results in the generalized second law of black hole thermodynamics.
 
 The paper is organized as follows: In the following section, we provide a brief review of the scalar-tensor theory of gravity. Thereafter, in the next section, we first describe the null-geometry and then obtain the DNS equation in the Jordan and Einstein frames. In the following section, we describe the evolution of the expansion scalar in the two frames by obtaining the Raychaudhuri equation in terms of the variables obtained earlier. There, we also prove the entropy increase theorem which follows from the obtained Raychaudhuri equation and from the null-energy condition. Then, in the penultimate section, we obtain the tidal force equation in each frame and show the evolution of the shear tensor. We end our work with the conclusions of our analysis.
\section{Brief review: equations of motion in the two frames}
Here, we briefly review the elements of ST gravity which we will use in this work. We begin by providing the relevant transformation relations between the Jordan and Einstein frames, and we introduce  the corresponding field equations of motion. These will be our main requirements since the fluid description of gravity comes from the projection of gravitational equation on to a null hypersurface. A more detailed analysis of scalar-tensor theory, at the level of the action, can be found in \cite{Bhattacharya:2017pqc, Bhattacharya:2018xlq}. For more about this action description and other important issues, one can look at our previous works \cite{Bhattacharya:2017pqc, Bhattacharya:2018xlq}. 

Usually, scalar-tensor theory is described in two frames. In the original frame, known as the Jordan frame, the action is given by
\begin{eqnarray}
&&\mathcal{A}=\int d^4x\sqrt{-g}L =\int d^4x\sqrt{-g} \frac{1}{16\pi G}\Big(\phi R
\nonumber
\\
&&-\frac{\omega (\phi)}{\phi}g^{ab}\nabla_a\phi \nabla_b\phi -V(\phi)\Big)~.
\label{SJ}
\end{eqnarray}
In this frame, the scalar field $\phi$ is non-minimally coupled to the Ricci-scalar $R$. Moreover $\omega (\phi)$, the Brans-Dicke parameter, is considered as a general function of $\phi$. When $\omega (\phi)$ is taken as a constant, the theory reduces to Brans-Dicke theory \cite{Faraoni:1999hp}. Here, $V(\phi)$ in \eqref{SJ} is an arbitrary scalar field potential. 
The equations of motion for $g_{ab}$ and $\phi$ are obtained by varying the action \eqref{SJ} yielding 
\begin{align}
E_{ab}=\frac{1}{16\pi G}[\phi G_{ab}+\frac{\omega}{2\phi}\nabla_i\phi\nabla^i\phi g_{ab}-\frac{\omega}{\phi}\nabla_a\phi\nabla_b\phi
\no 
\\
+\frac{V}{2}g_{ab}-\nabla_a\nabla_b\phi+g_{ab}\nabla_i\nabla^i\phi ]=0~; \label{EAB}
\end{align}
and
\begin{align}
E_{(\phi)}=\frac{1}{16\pi G}[R+\frac{1}{\phi}\frac{d\omega}{d\phi}\nabla_i\phi\nabla^i\phi +\frac{2\o}{\phi}\square\phi-\frac{dV}{d\phi}
\nonumber
\\
-\frac{\omega}{\phi^2}\nabla_a\phi \nabla^a\phi]=0~.
\label{EEAB}
\end{align}
$G_{ab}=R_{ab}-\frac{1}{2}Rg_{ab}$ is the Einstein tensor.  

The non-minimal coupling in \eqref{SJ} can be removed by a set of transformations: (i) a conformal transformation of the metric of the form
\begin{align}
g_{ab}\rightarrow\tilde{g}_{ab}=\Omega^2g_{ab},\ \ \ \ \ \ \ \ \Omega=\sqrt{\phi}~,
\label{GAB}
\end{align}
and (ii) a scaling of the scalar field given by 
\begin{align}
 \phi\rightarrow\tilde{\phi}\,\ {\textrm{with}}\,\ d\tilde{\phi}=\sqrt{\frac{2\omega+3}{16\pi G}}\frac{d\phi}{\phi}~.
\label{PHI}
\end{align}
From now on the untilde variables correspond to the Jordan frame while the tilde variable correspond to the Einstein frame.
Using the above relations \eqref{GAB} and \eqref{PHI}, the non-minimal coupling in \eqref{SJ} can be removed and one can go from the Jordan frame to the conformal (\ie Einstein frame). The action in the Einstein frame is
\begin{eqnarray}
&&\tilde{\mathcal{A}}=\int d^4x\sqrt{-\tilde{g}}\tilde{L}
\nonumber
\\
&&=\int d^4x\sqrt{-\tilde{g}}[\frac{\tilde{R}}{16\pi G}-\frac{1}{2}\tilde{g}^{ab}\tilde{\nabla}_a\tilde{\phi}\tilde{\nabla}_b\tilde{\phi}-U(\tilde{\phi})]~,
\label{SE}
\end{eqnarray}
where $U(\tilde{\phi}) = V(\phi)/(16\pi G\phi^2)$. The equations of motion of $\t g^{ab}$ and $\t\phi$ can be obtained from the variation of the action \eqref{SE}, which yields
\begin{align}
\t{E}_{ab}=\frac{\tilde{G}_{ab}}{16\pi G}-\frac{1}{2}\tilde{\nabla}_a\tilde{\phi}\tilde{\nabla}_b\tilde{\phi}+\frac{1}{4}\tilde{g}_{ab}\tilde{\nabla}^i\tilde{\phi}\tilde{\nabla}_i\tilde{\phi}
\nonumber
\\
+\frac{1}{2}\tilde{g}_{ab}U(\tilde{\phi})=0~; \label{EABTIL}
\end{align}
and 
\begin{align}
\t{E}_{(\t{\phi})}=\t{\na}_a\t{\nabla}^a\tilde{\phi}-\frac{dU}{d\tilde{\phi}}=0~.
\end{align}
One can check that under the transformations (\ref{GAB}) and (\ref{PHI}) the above equations reduces to Eqs. (\ref{EAB}) and (\ref{EEAB}). So both frames are equivalent, at the equations of motion level.  

We end this section by mentioning one key aspect not often highlighted in the literature. Although, the action in the two frames (\eqref{SJ} and \eqref{SE}) are connected by the set of transformations \eqref{GAB} and \eqref{PHI}, one can note that the actions in the two frames are equivalent only up to a total derivative term, which is the given as follows (for details see \cite{Bhattacharya:2017pqc, Bhattacharya:2018xlq}):
\begin{align}
\sqrt{-\tilde{g}}\tilde{L}=\sqrt{-g} L-\frac{3}{16\pi G}\sqrt{-g}\square\phi~.
\label{ACTUAL}
\end{align}
Since the two actions are connected by a total derivative term, it can be neglected if one is interested in the equations of motion \ie the dynamics of the system. In our previous work \cite{Bhattacharya:2017pqc, Bhattacharya:2018xlq}, we have shown that in constructing the thermodynamical quantities, like entropy, energy {\it etc.}, the extra total derivative term does play a crucial role.  This is because these boundary terms in the action do contribute to the thermodynamic parameters, since they are defined on the relevant boundary of the full manifold. We observed that the total derivative term indeed helps in resolving several inequivalences (such as thermodynamic entities and the holographic relations between the surface and bulk terms of the action) between the two frames. Of course, in the present case, the DNS equation and the dynamics of the null surface are obtained from the equations of motion of the metric tensor. Therefore, the extra $\square\phi$ term in \eqref{ACTUAL} has no significance in the present case. 

In the following section we will obtain the DNS equation for ST theory. In Einstein's gravity, when Einstein's equation is projected on to a null surface, it has a structure, similar to the Navier-Stokes equation of hydrodynamics \cite{DAMOUR}. In the present scenario with ST theory, we follow a similar approach. However, in this case, there are several extra terms in the equation of motion in the two frames. Also the scalar field $\phi$ is non-minimally coupled with the Ricci-scalar in the Jordan frame. Therefore, it will be interesting to investigate whether the DNS equation can be obtained in the present case of ST theory. If so, then we have to check how the expressions of the fluid parameters get modified. Also, we will examine whether the fluid parameters are conformally invariant between the two frames.



\section{DNS equation in Jordan and Einstein frames}
As mentioned above, the structure of the Navier-Stokes-like equation for GR was obtained first by Damour \cite{DAMOUR}. Later a more detailed investigation was performed by Price and Throne \cite{Price:1986yy}. After these early works various researchers expanded on this idea \cite{Gourgoulhon:2005ng, Padmanabhan:2010rp, Kolekar:2011gw, Parikh:1997ma, Bredberg:2011jq, Bhattacharyya:2008kq, Huang:2011he, Chirco:2011ex, Bai:2012ci, Bredberg:2010ky, Cai:2012mg, Zou:2013ix, Hu:2013lua, Cai:2011xv, Huang:2011kj, Anninos:2011zn, Ling:2013kua, Eling:2012ni, Berkeley:2012kz, Lysov:2017cmc, Wu:2013aov, Compere:2011dx, De:2019wok, De:2018zxo, Chatterjee:2010gp}. However, the approach which we follow here, depends largely on the structure of the null-hypersurface and the spacetime foliation of the null surface. Therefore, in the following we give a brief discussion of the null-hypersurface and its ($1+3$) foliation. More discussion in this regard, can be found in \cite{Gourgoulhon:2005ng}.

\subsection{Spacetime foliation on null hypersurface}

Let ($\mathcal{M}$, $g_{ab}$) be the whole ($1+3$)- dimensional manifold, where lower case Latin indices run $\{0, 1, 2, 3\}$. A null hypersurface (\H, $\g_{\a\b}$) is a three dimensional hypersurface embedded in a four-dimensional manifold \M such that it satisfies the condition $\g_{\a\b}v^{\b}=0$, where $v^{\a}$ is the vector defined in the tangent space of \H. Another way of saying this is that the pullback of the induced metric of a null-hypersurface onto its tangent plane is degenerate. Here, the Greek indices denote the coordinates associated to the null surface \H (defined by $x_3=$const.) and run $\{0, 1, 2\}$~. Moreover, null hypersurfaces are characterized by the null vectors $l^a$, which is orthogonal to \H and satisfies the geodesic condition of the spacetime. Since the normal of the null surface \H is self orthogonal, \ie $l^al_a=0$, one major difference of the null hypersurface \H with the spacelike or timelike one is that one cannot define an orthogonal projection operator onto \H. To study the dynamics of a null hypersurface extrinsically we also adopt the standard ($1+3$) foliation of the spacetime as described in \cite{Gourgoulhon:2005ng}.

In the present case, each spacelike hypersurface (which we denote as $\Sigma_t$, characterized by the unit timelike normal $\textbf{n}=N\textbf{d}t$ with $N$ being the lapse function) will intersect the null hypersurface \H on a two-dimensional surface $\mathcal{S}_t$ \ie $\mathcal{S}_t=\mathcal{H}\bigcap \Sigma_t$. Therefore, $\mathcal{S}_t$ will be characterized by two normals: one is the timelike normal $\textbf{n}$ and the other is a unit spacelike normal, denoted by $\textbf{s}$. These satisfy the properties $\textbf{s} \cdot \textbf{s}=1$, $\textbf{n} \cdot \textbf{n}=-1$ and $\textbf{n} \cdot \textbf{s}=0$~. Therefore, one can define an induced metric on the two surface $\mathcal{S}_t$, which is orthogonal to both $\textbf{n}$ and $\textbf{s}$~. The induced metric on $\mathcal{S}_t$ is defined as 
\begin{align}
q_{ab}=g_{ab}+n_an_b-s_as_b~. \label{INDUCED1}
\end{align}

Moreover, the two-surface $\mathcal{S}_t$ can also be regarded as the intersection of two null-hypersurfaces: one null surface $\mathcal{H}$, of course, is characterized by the outgoing null vector $\textbf{l}=N(\textbf{n}+\textbf{s})$ and the other one is characterized by the auxiliary (or ingoing) null vector $\textbf{k}=(1/2N)(\textbf{n}-\textbf{s})$, with the conditions $\textbf{l} \cdot \textbf{l}=\textbf{k} \cdot \textbf{k}=0$ and $\textbf{l}\cdot \textbf{k}=-1$~. The induced metric on the two surface $\mathcal{S}_t$, defined in \eqref{INDUCED1}, can now be written in terms of the null vectors $\textbf{l}$ and $\textbf{k}$ as
\begin{align}
q_{ab}=g_{ab}+l_ak_b+l_bk_a~. \label{INDJOR}
\end{align}
We shall use this form of the induced metric from now on. Note that $q_{ab}l^a=q_{ab}k^a=0$, and $q_{a}^bq^{c}_b=q_a^c$. Therefore, the induced metric $q_{ab}$ is orthogonal to the null vectors $\textbf{l}$ and $\textbf{k}$, and the mixed tensor $q^a_b$ allows one to project everything onto the two surface $\mathcal{S}_t$. 

This discussion and results are independent of what frame we are using, and are equally valid in both Einstein and Jordan frames. For example, 
the induced metric in the Einstein frame is identical to the form in \eqref{INDUCED1} and (\ref{INDJOR}) but with the quantities are denoted by the tilde variables: 
\begin{align}
\t q_{ab}=\t g_{ab}+\t n_a\t n_b-\t s_a\t s_b=\t g_{ab}+\t l_a\t k_b+\t l_b\t k_a~. \label{INDEIN}
\end{align}

Having all the necessary details, we are now ready to find the DNS-like equation by projecting the field equations. We begin the with the Einstein frame variables.

\subsection{DNS equation in the Einstein frame}
The procedure for obtaining the DNS equation in the Einstein frame is straightforward since the equation of motion in the Einstein frame (\ie Eq. \eqref{EABTIL}) is similar to the equation of motion from standard GR. 

The analysis starts by defining various important quantities on the null surface, which will ultimately be connected with the fluid variables.
We define $\t\th_{ab}$ as
\begin{align}
\t\th_{ab}=\t q^m_a\t q^n_b\t\na_m\t l_n=\t\na_a\t l_b+\t l_a\t k^i\t \na_i\t l_b-\t l_b\t\o_a~, \label{THETAABTIL}
\end{align}
which is the pullback of the covariant derivative of the null vector $\t{\textbf{l}}$ onto $\mathcal{S}_t$~. In the above relation \eqref{THETAABTIL}, we have defined 
\begin{equation}
\t \o_a=\t l^i\t\na_i\t k_a~.
\end{equation}
The expression of $\t\th_{ab}$ in \eqref{THETAABTIL} does not imply that $\t\th_{ab}$ is symmetric. However, if one replaces the term $\t l_a\t k^i\t\na_i\t l_b$ in \eqref{THETAABTIL} using the Frobenius theorem, \ie $\t l_{[a}\t\na_i\t l_{b]}=0$, one can straightforwardly obtain that $\t\th_{ab}$ is a symmetric tensor in $a$ and $b$~. $\t\th_{ab}$ can be decomposed into two parts: one being the symmetric trace-less part $\t\s_{ab}$ and the other is the trace part $\t\th$ given by
\begin{align}
\t\th_{ab}=\t\s_{ab}+\frac{\t q_{ab}}{2}\t\th~. \label{THETAAB1TIL}
\end{align}
The trace part $\t\th$ is given by 
\begin{align}
\t \th=\t q^{ab}\t\th_{ab}=\t\na_a\t l^a-\t\k~. \label{THETATIL}
\end{align}
In the above, $\t\kappa$ is the non-affinity of the null geodesics and is defined by the relation 
\begin{equation}
\t l^i\t\na_i\t l^a=\t\k\t l^a~.
\end{equation} 
Note that when a null-surface corresponds to a black hole horizon, $\t\kappa$ can be identified as the surface gravity of the black hole horizon.

Having the definition of different quantities on the null hypersurface, let us now project the equation (\ref{EABTIL}) on to this surface. Note that one part of this equation contains Einstein's tensor. The relevant part of this can be projected in the following way:
from $2\t\na_{[m}\t\na_{a]}\t l^m=\t R_{am}\t l^m$ (where, $A_{[i}B_{j]}\equiv(1/2)(A_iB_j-A_jB_i)$) and using \eqref{THETAABTIL} and \eqref{THETATIL} we obtain
\begin{align}
\t R_{am}\t l^m=\t\na_m\t\th^m_a+(\t\th+\t\k)\t\o_a+\t l^m\t\na_m\t\o_a-\t\na_a(\t\th+\t\k)
\no 
\\
-\t\na_m(\t l^a\t k^i\t\na_i\t l^m)~. \label{RAMLM1TIL}
\end{align}
Using \eqref{THETAABTIL}, one can further re-write the last term on the RHS of \eqref{RAMLM1TIL} as
\begin{align}
\t \na_m(\t l^a\t k^i\t \na_i\t l^m)=\t\th_{am}\t k^n\t\na_n\t l^m-\Big(\t\o_m\t k^n\t\na_n\t l^m
\no 
\\
-(\t\na_m\t k^i)(\t\na_i\t l^m)-\t k^i\t\na_m\t\na_i\t l^m\Big)\t l_a~. \label{LTTTTIL}
\end{align}
Substituting \eqref{LTTTTIL} in \eqref{RAMLM1TIL} we  obtain 
\begin{eqnarray}
&&\t R_{am}\t l^m=\t\na_m\t\th^m_a+(\t\th+\t\k)\t\o_a+\t l^m\t \na_m\t\o_a-\t\na_a(\t\th+\t\k)
\no 
\\
&&+\Big(\t\o_m\t k^n\t \na_n\t l^m-(\t\na_m\t k^i)(\t\na_i\t l^m)-\t k^i\t\na_m\t\na_i\t l^m\Big)\t l_a
\no 
\\
&&-\t\th_{am}\t k^n\t\na_n\t l^m~. \label{RAMLM2TIL}
\end{eqnarray}
Equation \eqref{RAMLM2TIL} plays a major role in the analysis. It will be shown later that when \eqref{RAMLM2TIL} is contracted with the projection operator $\t q^a_b$, it gives the DNS equation; when \eqref{RAMLM2TIL} is contracted with the null vector $\t l^a$, it results in the Raychaudhuri equation. 

We first obtain the DNS equation. Contracting (\ref{RAMLM2TIL}) with $\t q^a_b$ one obtains
\begin{align}
\t R_{mn}\t l^m\t q^n_a=\t q^n_a\t\na_m\t\th^m_n+(\t\th+\t \k)\t \O_a+\t q^n_a\t l^m\t \na_m\t\o_n
\no 
\\
-\t D_a(\t\th+\t\k)-\t\th_{am}\t k^i\t\na_i\t l^m~,\label{RMNLMQNATIL}
\end{align}
where $\t\O_a=\t q^b_a\t\o_b=\t\o_a+\t\k \t k_a$. In the above we used $\t l^a\t\o_a=\t\k$ and $\t k^a\t\o_a=0$. Also we denote $\t D_a(\t\th+\t\k)=\t q^b_a\t\na_b (\t\th+\t\k)$ where $\tilde{D}_a$ is the covariant derivative with respect the hypersurface's metric (\ie $\t q_{ab}$).
From equation \eqref{EABTIL} and using the identity \eqref{RMNLMQNATIL}, one obtains
\begin{align}
8\pi G\t T_{mn}^{(\t\phi)}\t l^m\t q^n_a=\t q^n_a\t\na_m\t\th^m_n-\t\th_{am}\t k^i\t\na_i\t l^m+\t q^n_a\t l^m\t\na_m\t\o_n
\no 
\\
+(\t\th+\t\k)\t\O_a-\t D_a(\t\th+\t\k)~, \label{TMNLMQNATIL}
\end{align}
where,
\begin{align}
\t T_{ab}^{(\t\phi)}=\tilde{\nabla}_a\tilde{\phi}\tilde{\nabla}_b\tilde{\phi}-\frac{1}{2}\tilde{g}_{ab}\tilde{\nabla}^i\tilde{\phi}\tilde{\nabla}_i\tilde{\phi}-\tilde{g}_{ab}U(\tilde{\phi})~. \label{TABTIL}
\end{align}
This can be identified as the energy-momentum tensor of the scalar-field $\t\phi$~. The first two terms on the right hand side (RHS) of \eqref{TMNLMQNATIL} can be replaced. First from the definition 
\begin{equation}
\t D_i\t\th^i_a=\t q^i_j\t q^k_a\t\na_i\t\th^j_k=\t q^k_a\t\na_i\t\th^i_k+(\t l^i\t k_j+\t l_j\t k^i)\t q^k_a\t\na_i\t\th^j_k~,
\end{equation} 
and using $\t\th^j_a\t l_j=\t\th^j_a\t k_j=0$, we obtain 
\begin{equation}
\t D_i\t\th^i_a=\t q^n_a\t\na_m\t\th^m_n-\t\th^j_a(\t l^i\t\na_i\t k_j+\t k^i\t\na_i\t l_j )~.
\end{equation} Therefore, we get
\begin{align}
\t D_i\t\th^i_a+\t\th_{am}\t\O^m=\t q^n_a\t\na_m\t\th^m_n-\t\th_{am}\t k^i\t\na_i\t l^m~. \label{REL1}
\end{align}
In addition, since $\t\o_i=\t\O_i-\t\k \t k_i$, one can obtain $\t q^n_a\t l^i\t\na_i\t\o_n=\t q^n_a\t l^i\t\na_i\t\O_n-\t\k\t\O_a$. Using the definition of the Lie-derivative, \ie $\tlie\t\O_n=\t l^i\t\na_i\t\O_n+\t\O_i\t\na_n\t l^i$, the previous expression can be further written  as 
\begin{align}
\t q^n_a\t l^i\t\na_i\t\o_n=\t q^n_a\tlie\t\O_n-\t\th^m_a\t\O_m-\t\k\t\O_a~. \label{REL2}
\end{align}
Substituting \eqref{REL1} and \eqref{REL2} in to \eqref{TMNLMQNATIL} one obtains
\begin{align}
8\pi G\t T_{mn}^{(\t\phi)}\t l^m\t q^n_a=\t q^n_a\tlie\t\O_n+\t\th\t\O_a-\t D_a(\frac{\t\th}{2}+\t\k)+\t D_i\t\sigma^i_a~, 
\label{TMNLMQNATIL1}
\end{align}
where, we have used $\t\th^m_n=\t\sigma^m_n+\frac{1}{2}\t q^m_n\t\th$~.
The RHS of equation \eqref{TMNLMQNATIL1} are the DNS terms in standard GR \cite{Gourgoulhon:2005ng, Padmanabhan:2010rp}. In this case our left hand side (LHS) is non-zero. Therefore \eqref{TMNLMQNATIL1}  is our DNS-like equation in the Einstein frame for ST gravity. It should be pointed out that, if the Lie-derivative in \eqref{TMNLMQNATIL1} is expressed in terms of the convective derivative, one extra term, $\t\th^i_a\t\O_i$, appears compared to the usual Navier-Stokes (NS) equation. This term does not have any fluid dynamic interpretation. This difference was highlighted in earlier works \cite{Price:1986yy, Gourgoulhon:2005ng, Padmanabhan:2010rp}.

 The corresponding fluid parameters of the DNS equation in the Einstein frame as given in \eqref{TMNLMQNATIL1} are identified as follows: the term on LHS, $\t F_{a}=\t T_{ab}^{(\t\phi)}\t l^b$, can be identified as the external force term; the momentum density is given as $\t\pi_a=-\t\O_a/8\pi G$; the pressure is identified as $\t P=\t\k/8\pi G$; the shear viscosity coefficient is given as $\t\eta=1/16\pi G$; the bulk viscosity coefficient is given as $\t\xi=-1/16\pi G$ (note that for the standard NS equation, the total viscous tensor is $2\eta\s^a_b+\xi\d^a_b\th$)~.

 In the present case, one does not necessarily require any external matter source to have an $\t F_{a}$ term in the DNS equation \eqref{TMNLMQNATIL1}. This is the major difference with standard GR. If one includes some external matter, one can still obtain the DNS equation. However in this case the energy-momentum tensor $\t T_{ab}^{(\t\phi)}$ in \eqref{TMNLMQNATIL1} will be replaced by $\t T_{ab}=\t T_{ab}^{(\t\phi)}+\t T_{ab}^{(ext)}$, where $\t T_{ab}^{(ext)}$ is the energy-momentum tensor coming from the additional external matter source. The external force in the DNS equation is then identified as $\t F_{a}=\t T_{ab}\t l^b$.
 
 We make one final comment before moving on to the Jordan frame. The entropy of scalar-tensor theory in the Einstein frame is identified as \cite{Koga:1998un, Bhattacharya:2017pqc, Bhattacharya:2018xlq} $\t S=\t A/4G$, where $\t A$ is the surface area of the null horizon. This gives an entropy density of $\t s=\t S/\t A=1/4G$~. As a result, the ratio of shear viscosity to entropy density is 
 \begin{align}
 \frac{\t\eta}{\t s}=\frac{1}{4\pi}~. \label{KSSEIN}
 \end{align}
 This is the same as in standard GR and is consistent with the Kovtun-Son-Starinets (KSS) bound \cite{Kovtun:2004de}.
 \subsection{DNS equation in Jordan frame}
In obtaining the DNS equation in the Jordan frame, there are several points one has to recognize. First, the scalar field in the Jordan frame is non-minimally coupled with the Ricci scalar and the gravitational interaction is mediated both by the metric tensor as well as by the scalar field. Therefore, it is expected that the scalar field will invariably appear in the expressions of the fluid parameters and the transport coefficients. Moreover, we mentioned earlier that both the equations of motion for the metric (\ie (\ref{EAB}) and (\ref{EABTIL})) in the two frames are equivalent. Thus, a proper DNS equation in the Jordan frame will be one where the scalar field will appear in the expressions of the fluid parameters as well as having the equation be conformally connected to (\ref{TMNLMQNATIL1}). Note that the DNS equation and the fluid parameters in the Einstein frame have been unambiguously obtained at the end of the last subsection.
 However, in the Jordan frame, the presence of $\phi$ gives rise to two different pictures/cases in the fluid description. In the first picture (\ie case 1), the fluid variables are not conformally equivalent in the two frames. Further, in case 1, the fluid dynamic equation appears as the DNS equation along with an extra Coriolis-like force term. Case 1, as we discuss in the following, agrees with some works, which are in the context of the AdS/CFT correspondence. However we provide thorough discussions about the relevance of these earlier works in the present context. In the second picture (case 2) one finds that the fluid variables in the two frames are conformally equivalent. In spite of the fact that this picture may violate the KSS bound for some range of $\phi$, we discuss how this picture agrees with the classical viewpoint. In the subsections below we first discuss the inequivalent picture (case 1), and then we move on to discuss the equivalent picture (case 2).

\subsubsection{Inequivalent picture -- case 1}

We follow a procedure similar to that used above in the Einstein frame, to obtain various useful results, that help us to obtain the final form of the DNS equation in the Jordan frame. First we find $\th_{ab}$ as
\begin{align}
\th_{ab}=q^m_aq^n_b\na_ml_n=\na_al_b+l_ak^i\na_il_b-l_b\o_a~, \label{THETAAB}
\end{align}
with $\o_a= l^i \na_i k_a$.
As before, in the Einstein frame case, $\th_{ab}$ can be written as sum of a traceless part and a trace part:
\begin{align}
\th_{ab}=\s_{ab}+\frac{q_{ab}}{2}\th~, \label{THETAAB1}
\end{align}
where, $\th$ is given as 
\begin{align}
\th=q^{ab}\th_{ab}=\na_al^a-\k~. \label{THETA}
\end{align}
In addition, we find
\begin{eqnarray}
&&R_{am}l^m=\na_m\th^m_a+(\th+\k)\o_a+l^m\na_m\o_a-\na_a(\th+\k)
\no 
\\
&&+\Big(\o_mk^n\na_nl^m-(\na_mk^i)(\na_il^m)-k^i\na_m\na_il^m\Big)l_a
\no 
\\
&&-\th_{am}k^n\na_nl^m~. \label{RAMLM2}
\end{eqnarray}
Projecting the above expression onto the null surface, we obtain
\begin{align}
R_{mn}l^mq^n_a=q^n_a\na_m\th^m_n+(\th+\k)\O_a+q^n_al^m\na_m\o_n
\no 
\\
-D_a(\th+\k)-\th_{am}k^i\na_il^m~,\label{RMNLMQNA}
\end{align}
where $\O_a= q^b_a\o_b=\o_a+\k k_a$ with $ l^a \o_a=\k$; $ k^a \o_a=0$ and $ D_a(\th+\k)=  q^b_a\na_b (\th+\k)$~. As in the Einstein frame, we can obtain two additional identities
\begin{align}
 D_i\th^i_a+\th_{am}\O^m= q^n_a\na_m\th^m_n-\th_{am} k^i\na_i l^m~. \label{REL1JR}
\end{align}
and 
\begin{align}
 q^n_a l^i\na_i\o_n= q^n_a\lie\O_n-\th^m_a\O_m-\k\O_a~. \label{REL2JR}
\end{align}
Substituting \eqref{REL1JR} and \eqref{REL2JR} in \eqref{RMNLMQNA} one obtains
\begin{align}
R_{mn} l^m q^n_a= q^n_a\lie\O_n+\th\O_a- D_a(\frac{\th}{2}+\k)+ D_i\sigma^i_a~. \label{NOTDNS1}
\end{align}

The LHS of \eqref{NOTDNS1} can be written as $R_{mn} l^m q^n_a=8\pi G\mathcal{T}_{mn} l^m q^n_a$ where, using the equation of motion \eqref{EAB}, one can identify $\mathcal{T}_{mn}$ as
\begin{align}
\mathcal{T}_{mn}=\frac{\o}{8\pi G\phi^2}\Big(\na_m\phi\na_n\phi-\frac{1}{2} g_{mn}\na^i\phi\na_i\phi\Big)-\frac{V g_{mn}}{16\pi G\phi}
\no 
\\
+\frac{1}{8\pi G\phi}\Big(\na_m\na_n\phi- g_{mn}\na_i\na^i\phi\Big)~.
\end{align}
Therefore, one obtains
\begin{align}
8\pi G\mathcal{T}_{mn} l^m q^n_a= q^n_a\lie\O_n+\th\O_a- D_a(\frac{\th}{2}+\k)+ D_i\sigma^i_a~. \label{NOTDNS}
\end{align}
Although equation \eqref{NOTDNS} looks like the gravitational Navier-Stokes equation for the Jordan frame there are issues with making this identification.  In the above relation \eqref{NOTDNS}, the scalar field $\phi$ does not appear in the expressions of the fluid parameters and the transport coefficients. The role of the scalar field in \eqref{NOTDNS} appears like an external field, which is not consistent with the non-minimal coupling in ST theory in the Jordan frame. Therefore, we conclude that Eq. \eqref{NOTDNS} is not the correct form of the DNS equation in the Jordan frame. 

 To make Eq.\eqref{NOTDNS}  consistent with the non-minimal coupling, we incorporate $\phi$ in the expressions of the fluid parameters and the transport coefficients, to obtain a proper DNS equation in the Jordan frame. We begin by multiplying \eqref{NOTDNS1} by $\phi$, and using some identities to move $\phi$ past some of the derivative operators.
\begin{align}
\phi R_{mn} l^m q^n_a= q^n_a\lie(\phi\O_n)+\th(\phi\O_a)- \frac{\phi}{2} D_a\th
\no 
\\
-D_a(\phi\k)+\phi D_i\sigma^i_a+2l^m q^n_a \Big(\omega_{[m}\na_{n]}\phi\Big)~. \label{ACTDNS1}
\end{align}
Using the equation of motion \eqref{EAB}, equation \eqref{ACTDNS1} can be written as
\begin{align}
8\pi G T_{mn}^{(\phi)} l^m q^n_a= q^n_a\lie(\phi\O_n)+\th(\phi\O_a)- \frac{\phi}{2} D_a\th
\no 
\\
-D_a(\phi\k)+\phi D_i\sigma^i_a+2l^m q^n_a \Big(\omega_{[m}\na_{n]}\phi\Big)~, \label{ACTDNSFINAL}
\end{align}
where $T_{mn}^{(\phi)}=\phi \mathcal{T}_{mn}$. The RHS of the above equation, other than the last term, has the structure of the DNS equation. 
The last term, $2l^m q^n_a\Big(\omega_{[m}\na_{n]}\phi\Big)$, can be identified as the rotational contribution in the full manifold. This term can be thought of as the rotational contribution in the DNS equation and therefore we identify it as a Coriolis-like force term. The expression of the DNS equation in \eqref{ACTDNSFINAL} is the counterpart of (\ref{TMNLMQNATIL1}) in the Einstein frame, as can be checked by applying the transformations (\ref{GAB}) and (\ref{PHI}). 

We identify the fluid parameters and the transport coefficient in the following way:
the external force is identified as $F_a= T^{(\phi)}_{ab}l^b$; the momentum density is identified as $\pi_a=-\phi\O_a/8\pi G$; the pressure is given as $P=\phi\k/8\pi G$; the shear viscosity coefficient is identified as $\eta=\phi/16\pi G$; the bulk  viscosity coefficient as $\xi=-\phi/16\pi G$. Thus, equation \eqref{ACTDNSFINAL} can be interpreted as the hydrodynamic equation of a fluid in a rotating frame with the angular velocity $W_a=\na_a\phi/2$.

Even without an external matter field, we get an external force term coming from the $T^{(\phi)}_{ab}$ term. When an external matter term is added the external force term is given by the combination of the energy-momentum tensor for $\phi$ plus the energy-momentum tensor of the external field $\ie$  $T_{ab}=T^{(\phi)}_{ab}+T_{ab}^{ext}$. As in previous cases the external force will be identified as $F_a= T_{ab}l^b$.
Note that, in this case, the ratio of the shear viscosity $\eta$ (which here is equal to $\phi/16\pi G$) to the entropy density is $\eta/s=1/4\pi$. This value is consistent with KSS bound \cite{Kovtun:2004de}. A similar $\eta/s$ ratio has also been predicted in \cite{Brustein:2008cg} for $f(R)$ gravity, which can be considered as a subclass of scalar-tensor gravity. Therefore, the inequivalent picture, which has been described here,   agrees with the literature. Note that the scalar field $\phi$, which is non-minimally coupled in this frame, appears in several fluid parameters (like $\pi_a$, $P$, $\eta$, $\xi$). One way of interpreting the non-minimal coupling of $\phi$ with $R$ is that in the Jordan frame, Newton's constant of GR is replaced by an effective Newton's constant as $G\longrightarrow G_{eff}=G/\phi$~\cite{Faraoni:2010yi}. With this viewpoint, the modification of the expressions of fluid parameters and the transport coefficients can be justified and those expressions can be seen as being the same as in GR except with $G$ rescaled as $G/\phi$.
 
 Although the above case agrees with earlier theoretical results in the literature, it shows that the fluid variables in the two frames are not equivalent ({\it e.g.} $\t P\neq P$, $\t\pi_a\neq\pi_a$, $\t F_a\neq F_a$ {\it etc.}). We now clarify this inequivalence of the fluid variables.
 From the conformal transformation relation \eqref{GAB} and from the normalization conditions of $\textbf{n}$ and $\t{\textbf{n}}$ (\ie $\textbf{n}\cdot \textbf{n}=\t{\textbf{n}} \cdot \t{\textbf{n}}=-1$) it follows that $\t n_a=\sqrt{\phi}n_a$ and $\t n^a=(1/\sqrt{\phi})n^a$~. Similarly from the normalization condition of $\textbf{s}$ and $\t{\textbf{s}}$ one finds $\t s_a=\sqrt{\phi}s_a$ and $\t s^a=(1/\sqrt{\phi})s^a$~. Moreover, in the ($1+3$) foliation of the spacetime, the lapse function $N$ is given as $N=\sqrt{-g_{00}}$ so we find $\t N=\sqrt{\phi}N$. Earlier we defined $\textbf{l}$ and $\textbf{k}$ as linear combinations of $\textbf{n}$ and $\textbf{s}$, which yields 
\begin{align}
\t l^a=l^a, \ \ \ \ \ \  \t l_a=\phi l_a
\no 
\\
\t k^a=\frac{1}{\phi} k^a, \ \ \ \ \ \  \t k_a= k_a~. \label{LNK}
\end{align}
The transformation relation of various quantities, that define the fluid variables in the Jordan frame and the Einstein frame, are given by
\begin{eqnarray}
&&\t\th_{a}^b=\th_{a}^b+\frac{q_{a}^b}{2}l^i\na_i\ln\phi~,
\no 
\\
&&\t\th=\th+l^i\na_i\ln\phi~,
\no 
\\
&&\t\sigma^i_j=\sigma^i_j~,
\no 
\\
&&\t\k=\k+l^i\na_i\ln\phi~,
\no 
\\
&&\t\o_a=\o_a+\frac{1}{2}[l_ak^i\na_i\ln\phi+\na_a\ln\phi-k_al^i\na_i\ln\phi]~,
\no 
\\
&&\t\O_a=\O_a+\frac{1}{2}q^b_a\na_b\ln\phi~.
 \label{TRANSREL}
\end{eqnarray}
From the relations in \eqref{TRANSREL}, one can conclude that the fluid variables in the two frames are not equivalent for the present case. As mentioned earlier, whether the two conformally connected frames are physically equivalent or not, has been an unsolved puzzle for several decades. However, from the thermodynamic perspective, we have shown that the two frames are thermodynamically equivalent and the thermodynamic parameters are exactly equivalent between the two frames \cite{Bhattacharya:2017pqc, Bhattacharya:2018xlq} (under certain assumptions, similar results have also been obtained in \cite{Koga:1998un}). All of this leads to the question ``Is it possible to obtain a DNS-like equation in the Jordan frame, where the fluid variables are equivalent across the two frames?". In the next subsection we show that one can answer this question in the affirmative. 

\subsubsection{Equivalent picture -- case 2}
We now want to alter our previous analysis of the Jordan frame to find a DNS equation whose fluid variables are the same as those in the Einstein frame (\ie $\t\k$, $\t\th$, $\t\sigma^a_b$, $\t \O_a$ {\it etc.}). We start by examining the expression: $q^n_a\na_m\t\th^m_n+(\t\th+\t\k)\t\O_a+q^n_al^m\na_m\t\o_n-q^n_a\na_n(\t\th+\t\k)-\t\th^m_ak^i\na_il_m$~. In this expression the fluid variables are taken intentionally as that of the Einstein frame ({\it e.g.} $\t\o_a$, $\t\k$, $\t\theta$, $\t\theta_{ab}$, $\t\O_a$ {\it etc.}), but the background and the covariant derivatives are defined with respect to the metric of the Jordan frame. Using \eqref{TRANSREL}, we obtain
\begin{eqnarray}
&&q^n_a\na_m\t\th^m_n+(\t\th+\t\k)\t\O_a+q^n_al^m\na_m\t\o_n-q^n_a\na_n(\t\th+\t\k)
\no 
\\
&&-\t\th^m_ak^i\na_il_m=R_{mn}l^mq^n_a-q^n_al^i\na_n\na_i\ln\phi
\no 
\\
&&+2l^i\O_a(\na_i\ln\phi) -\frac{3}{2}q^n_a(\na_nl^i)(\na_i\ln\phi)
\no 
\\
&& +\frac{1}{2}(\k+\th)q^b_a\na_b\ln\phi + q^b_al^i(\na_i\ln\phi)(\na_b\ln\phi)~.
 \label{RELJOR1}
\end{eqnarray}
To obtain this result, we have used  \eqref{RMNLMQNA}.
Now, one can straightforwardly obtain the following relation
\begin{equation}
D_b\t\th^b_a=q^i_jq^k_a\na_i\t\th^j_k=q^n_a\na_m\t\th^m_n-\t\th^j_a\o_j-\t\th^m_ak^i\na_il_m~.
\end{equation}  
From this expression and using \eqref{TRANSREL}, one further finds 
\begin{align}
q^n_a\na_m\t\th^m_n-\t\th^m_ak^i\na_il_m=D_b\t\th^b_a+\th^j_a\O_j+\frac{\O_a}{2}l^k\na_k\ln\phi~. \label{RELJOR2}
\end{align}
Also we have, $q^n_al^m\na_m\t\o_n=q^n_al^m\na_m(\t\O_n-\t\k\t k_n)=q^n_a\lie\t\O_n-\th^m_a\t\O_m-\O_a\t\k$~. Again, using the transformation relation \eqref{TRANSREL}, we arrive at
\begin{align}
q^n_al^m\na_m\t\o_n=q^n_a\lie\t\O_n-\th^j_a\O_j-\t\O_a\t\k-\frac{\th^i_a}{2}\na_i\ln\phi
\no 
\\
+\frac{\k}{2}q^b_a(\na_b\ln\phi)+\frac{q^b_a}{2}l^i(\na_b\ln\phi)(\na_i\ln\phi)~.\label{RELJOR3}
\end{align}
We now substitute \eqref{RELJOR2} and \eqref{RELJOR3} into \eqref{RELJOR1} which yields
\begin{align}
D_b\t\th^b_a+q^n_a\lie\t\O_n-\frac{\th^i_a}{2}\na_i\ln\phi+\t\th\t\O_a-D_a(\t\th+\t\k)=
\no 
\\ 
\Big(R_{mn}-\na_m\na_n\ln\phi+\frac{1}{2}(\na_m\ln\phi)(\na_n\ln\phi)\Big)l^mq^n_a
\no 
\\
-\frac{3}{2}q^n_a(\na_nl^i)(\na_i\ln\phi)+\frac{3}{2}\O_al^i\na_i\ln\phi+\frac{\th}{2}q^b_a(\na_b\ln\phi)~. \label{RELJOR4}
\end{align}
Equation \eqref{RELJOR4} can be further simplified. From \eqref{THETAAB}, one obtains $q^n_a(\na_nl^i)(\na_i\ln\phi)=\O_al^i\na_i\ln\phi+\th^i_a(\na_i\ln\phi)$~. Using this with $\th^i_a=\sigma^i_a+\frac{q^i_a}{2}\th$ in \eqref{RELJOR4}, yields
\begin{align}
q^n_a\lie\t\O_n-D_a(\frac{\t\th}{2}+\t\k)+\t\th\t\O_a+D_b\t\sigma^b_a+\t\sigma^i_a(\na_i\ln\phi)
\no 
\\
=\Big(R_{mn}-\na_m\na_n\ln\phi+\frac{1}{2}(\na_m\ln\phi)(\na_n\ln\phi)\Big)l^mq^n_a~. \label{RELJOR5}
\end{align}
Equation \eqref{RELJOR5} can also be identified as a DNS-like equation in the Jordan frame. From the equation of motion \eqref{EAB}, we obtain
\begin{eqnarray}
&&R_{mn}-\na_m\na_n\ln\phi+\frac{1}{2}(\na_m\ln\phi)(\na_n\ln\phi)=
\no 
\\
&&\frac{(2\o+3)}{2}(\na_m\ln\phi)(\na_n\ln\phi)+g_{mn}\Big(\frac{R}{2}
\no 
\\
&&-\frac{1}{2}(\na_i\phi)(\na^i\phi)-\frac{V}{2\phi}-\frac{1}{\phi}\square\phi\Big)~.\label{RELJOR6}
\end{eqnarray}
Also, using the transformation relations in \eqref{TABTIL} one can obtain an expression for $\t T_{ab}^{(\t\phi)}$ in the Jordan frame as
\begin{align}
\t T_{ab}^{(\t\phi)}=\frac{(2\o+3)}{16\pi G}\Big[(\na_a\ln\phi)(\na_b\ln\phi)-\frac{1}{2}g_{ab}(\na_i\phi)(\na^i\phi)\Big]
\no 
\\
+\frac{V}{16\pi G\phi}g_{ab}~.\label{RELJOR7}
\end{align}
Comparing equations \eqref{RELJOR6} and \eqref{RELJOR7}, one can conclude that in \eqref{RELJOR5} the RHS, as a whole, contributes to $8\pi G\t T_{mn}^{(\t\phi)} l^m q^n_a$~.
Thus the expression of the DNS-like equation \eqref{RELJOR5} in the Jordan frame, is given as
\begin{align}
q^n_a\lie\t\O_n-D_a(\frac{\t\th}{2}+\t\k)+\t\th\t\O_a+\frac{1}{\phi} D_b\Sigma^b_a=8\pi G\t T_{mn}^{(\t\phi)} l^m q^n_a~.\label{DNSJOr}
\end{align}
One can again identify \eqref{DNSJOr} as the DNS equation in the Jordan frame, where the fluid variables are equivalent to those in the Einstein frame and are identified as follows: the external force is identified as $F_a=\t T^{\t\phi}_{ab}l^b$; the momentum density is identified as $\pi_a=-\t\O_a/8\pi G$; the pressure is given as $P=\t\k/8\pi G$; shear tensor is $\Sigma^a_b$; the shear viscosity coefficient is identified as $\eta=1/(16\pi G\phi)$; the bulk  viscosity coefficient is $\xi=-1/16\pi G$~.

For this case the DNS equation in the Jordan frame, equation \eqref{DNSJOr}, all the fluid parameters are equivalent to those in the Einstein frame (\ie $\t F_{a}=F_{a}, \t\pi_a=\pi_a$... {\it etc.}) except for the shear viscosity coefficient, which is connected as $\eta=\t\eta/\phi$. Also the shear tensor in the Jordan frame for this case is connected to the the shear tensor in the Einstein frame as $\Sigma^a_b=\phi\t\sigma^a_b$ and $\Sigma_{ab}=\t\sigma_{ab}$~. 

We now examine what happens when an external matter field is included in the gravitational action. In the Jordan frame, the action of the external matter field is $\mathcal{A}_{ext}=\int\sqrt{-g}\mathcal{L}_{ext}d^4x$, whereas in the Einstein frame, it is given as $\mathcal{\t A}_{ext}=\int\sqrt{-\t g}\mathcal{\t L}_{ext}d^4x$~. Since these actions for the external fields in the two frames are conformally invariant (\ie $\mathcal{\t A}_{ext}=\mathcal{A}_{ext}$), this implies that $\sqrt{-\t g}\mathcal{\t L}_{ext}=\sqrt{-g}\mathcal{L}_{ext} $ and $\mathcal{\t L}_{ext}=\mathcal{L}_{ext}/\phi^2$~. Now, in the Jordan frame, the energy momentum tensor corresponding to the external matter source is
\begin{align}
T_{ab}^{(ext)}=-\frac{2}{\sqrt{-g}}\frac{\d(\sqrt{-g}\mathcal{L}_{ext})}{\d g^{ab}}~,
\end{align}
which is connected to the energy-momentum tensor in the Einstein frame as 
\begin{align}
\label{EMT-JE}
\t T_{ab}^{(ext)}=-\frac{2}{\sqrt{-\t g}}\frac{\d(\sqrt{-\t g}\mathcal{\t L}_{ext})}{\d \t g^{ab}}=\frac{T_{ab}^{(ext)}}{\phi}~.
\end{align}
Introducing an external matter field results in the zero on the RHS of \eqref{EAB} being replaced by $T^{(ext)}_{ab}/2$. However, in \eqref{RELJOR6}, one gets the extra term $8\pi G T^{(ext)}_{ab}/\phi$~.
Thus, taking the external matter field into account, gives the following result for the external force in the Jordan frame: $F_a=(\t T^{\t\phi}_{ab}+(T_{ab}^{(ext)}/\phi))l^b=(\t T^{\t\phi}_{ab}+\t T_{ab}^{(ext)})l^b=\t F_a$~. In summary the introduction of an external matter field leads to a force that is invariant between the Jordon and Einstein frames, while the energy-momentum tensors are related by a factor of $1/\phi$ as seen in \eqref{EMT-JE}.  
 
 
 We conclude this section by making a final important comment. The fluid-gravity connection is not well-explored in the context of the scalar-tensor gravity. However, there are a few works for $f(R)$ gravity \cite{Brustein:2008cg, Chatterjee:2010gp}, which suggest that the shear viscosity coefficient is $\eta=f'(R)/16\pi G$ which lead to the following ratio with the entropy density $\eta/s=1/4\pi$. This ratio saturates the KSS bound. Since $f(R)$ gravity can be analyzed as a particular form of scalar-tensor gravity with the scalar field identified as $\phi\equiv f'(R)$, those works on $f(R)$ gravity favor the inequivalent picture (case 1) of the Jordan frame described in the last subsection. But in the present picture (case 2) we have
 \begin{equation}
 \label{eta-2}
 \frac{\eta}{s} = \frac{1}{4\pi \phi^2}~,
 \end{equation}
 which is not a constant and depends on $\phi$. The
 ratio in \eqref{eta-2} can violate the KSS bound, $ \frac{\eta}{s} \ge \frac{1}{4 \pi}$, if $\phi$ is greater than 1. Let us now discuss this issue in a more detail.
 
 As mentioned earlier, using the AdS/CFT correspondence, it is argued in the literature (for example in \cite{Brustein:2008cg}) that the ratio $\eta/s$ is the same for any pair of gravity theories which are connected with each other by a field redefinition. Furthermore, it has been noted \cite{Brustein:2008cg} that if a gravity theory (such as $f(R)$ gravity), after a field redefinition, can be written as Einstein's GR along with the terms for matter fields, its $\eta/s$ ratio is always $1/4\pi$ (i.e. the value of Einstein's GR). The presence of external matter fields does not modify $\eta/s$ unless the matter is coupled with the Riemann tensor.  The argument, proposed in \cite{Brustein:2008cg}, also works for other gravity theories \cite{Brigante:2007nu, Brigante:2008gz, Kats:2007mq, Son:2007vk, Buchel:2004di}. For example, according to \cite{Brigante:2007nu}, $\eta/s$ violates the KSS bound (see Eq. 1.3 of this reference) for a higher order derivative theory, containing Einstein-Gauss-Bonnet gravity terms with different multiplicative coefficients in front of each term. However, when one of the coefficients ($\alpha_3$ of Ref. \cite{Brigante:2007nu}) is set to vanish, the theory can be redefined (see Eq. 2.9 of \cite{Brigante:2007nu}) as the Einstein's gravity. In this case, $\eta/s$ (obtained from AdS/CFT) coincides with the value of Einstein's gravity. The same analysis has also been done in the appendix A of \cite{Kats:2007mq} for higher derivative gravity theories. Therefore, the proposal provided in \cite{Brustein:2008cg}, is also consistent with higher derivative gravity theories. In addition, the arguments of \cite{Brustein:2008cg} also work for type IIB supergravity in ten spacetime dimensions \cite{Son:2007vk}, which can be reduced to five-dimensional Einstein's theory with the massless dilaton, SO(6) gauge bosons, and gravitons, by performing a Kaluza-Klein (KK) reduction. In this case, one obtains $\eta/s=1/4\pi$ from the direct calculation using AdS/CFT in ten dimensions \cite{Son:2007vk}. In a similar context, it has been found \cite{Buchel:2004di}, that the low-energy effective action of type IIB string theory in ten dimensions along with the leading order string corrections, cannot be reduced to Einstein's theory \cite{Son:2007vk} as the matter remains coupled with the Riemann tensor even after the KK reduction. In this case, $\eta/s$ is not only $1/4\pi$, but is modified by the coupling term \cite{Buchel:2004di}.

 Scalar-tensor theory in the Jordan frame can be written as Einstein's gravity along with scalar field terms (which are not coupled with the Riemann tensor) in the Einstein frame. Moreover, as mentioned earlier, AdS/CFT computations have been done in \cite{Brustein:2008cg} for $f(R)$ gravity, which is a subclass of the general scalar-tensor theory. For $f(R)$ gravity, $\eta/s$ leads to $\eta/s=1/4\pi$. Therefore, as per the argument in \cite{Brustein:2008cg}, the direct computation of $\eta/s$ using AdS/CFT should result in $1/4\pi$ both in the Einstein frame as well as in the Jordan frame for the case of scalar-tensor theory. In the present paper we found this ratio using a completely different approach, i.e. via the fluid-gravity correspondence. In the in-equivalent picture (i.e. case 1) we observed that $\eta/s=1/4\pi$. Therefore this observation again bolsters the proposal made in \cite{Brustein:2008cg}, but here at the cost of the in-equivalence of different fluid parameters. In this regard it is to be noted that the fluid parameters are obtained from the derivative of the null-like vectors (projected on the two-surface). These vectors are connected by the relation (\ref{LNK}) in the two frames and these are not equivalent. Therefore, it may also be possible that the fluid parameters are not equivalent (which is a subject for further investigation) in the two frames. Hence, in that sense, our in-equivalent picture with saturation of the KSS bound is apparently consistent with the already existing proposal in literature.
 
However, it is well known that, the KSS bound actually originates in the specific context of AdS/CFT correspondence. On the other hand, the fluid-gravity correspondence also provides the value of $\eta$ and gives a ratio $\eta/s$ (similar to what has been done in the present paper) which matches the $\eta/s$ ratio obtained in the context of AdS/CFT in Einstein's gravity. But, whether these two pictures must always be compatible with each other, has yet to be shown concretely. Therefore, the same may not happen in scalar-tensor gravity. This issue needs to be investigated further. For this reason, the $\eta/s$ ratio has to be obtained using the AdS/CFT correspondence and it has to be checked whether the result obtained from AdS/CFT matches the result obtained from our fluid-gravity analysis in \textit{any of the two pictures}. However, to formulate a consistent AdS/CFT correspondence in scalar-tensor gravity, one needs to check whether an asymptotic AdS solution exists in this theory and also whether one can formulate a causally-well-behaved AdS/CFT correspondence using that solution. Therefore, further investigation is required in this direction to formulate a consistent AdS/CFT correspondence for scalar-tensor gravity.
  
On the other hand, since the two frames are classically equivalent, we formulate ``case 2'' in the same spirit, considering the fluid parameters to be equivalent in the two frames. In that process, we found that $\eta/s$ can violate the KSS bound when $\phi>1$. We do not undervalue  ``case 2'' relative to ``case 1" for that reason, since as discussed above the result for the ratio, $\eta/s$, from the fluid/gravity analysis might not agree with the result obtained from AdS/CFT for alternative gravity theories, such as the scalar-tensor theories. Moreover, from other arguments (e.g. forbidding naked singularities in the spacetime solution, maintaining causality etc.) we also have to check whether values with $\phi>1$, which violated the KSS bound, are allowed or not. For example, it has been found in Gauss-Bonnet gravity that the KSS bound is violated for positive values of the coupling constant $\lambda_{GB}$ \cite{Brigante:2007nu}. However, it has been found that the value of $\lambda_{GB}$ has to be $\lambda_{GB}\leq 1/4$ to avoid a naked singularity in the solution. Further analysis suggests that $\lambda_{GB}$ has to be $\lambda_{GB}\leq 9/100$ to maintain causality \cite{Brigante:2007nu} (also see \cite{Camanho:2014apa}, which argues that the causality is violated for the coupling constant $\gg l_P^2$). So within the allowed range of $\lambda_{GB}$ ($\lambda_{GB}\leq 9/100$) for which the solution in Einstein-Gauss-Bonnet gravity exists and the causality is maintained, the KSS bound may be violated (for $0<\lambda_{GB}\leq 9/100$). Thus for particular values of $\phi$, the violation of the KSS bound in our scalar-tensor theory for ``case 2'' can occur, provided those values of $\phi$ are consistent with the existence of the solutions and also with the causality of this theory. This has to be checked using conditions like no violation of causality, as done in \cite{Camanho:2014apa, Papallo:2015rna} for different gravity theories.

It is clear that further investigation is required to check whether the fluid parameters are indeed equivalent in the two frames.  Finding the DNS equation on the null surface, for scalar-tensor theory, indeed shows the possibility of fluid-gravity correspondence in this theory. But as of now, it does not concretely provide any insight on whether the fluid parameters need to be conformally invariant. This can be indirectly investigated through the AdS/CFT formalism using AdS solutions of this theory. Once the KSS ratio is obtained in this way, one may use the  value of the KSS ratio to single out one of our present proposed pictures, which will be consistent with the AdS/CFT analysis. This task needs a completely independent investigation. We keep it as an open issue and leave it for future work. For now our analysis does not rule out either ``case 1'' or ``case 2'', but rather provides further motivation to work along this direction. One has to check from other analysis which of these pictures is more consistent.
\section{The entropy increase theorem}
Within the framework of standard GR, when $R_{mn}l^n$ is projected onto the null surface by contraction with the projection tensor, $q^m_a$, it yields the DNS equation. On the other hand, when $R_{mn}l^n$ is contracted with another null vector, $l^m$, it results in the Raychaudhuri equation. In an earlier section, we showed that the procedure of obtaining the DNS equation in scalar-tensor theory (in Jordan frame) is different from standard GR due to the scalar field $\phi$. In this section, we discuss the Raychaudhuri equation in the context of the scalar-tensor gravity. Moreover, it is well-known that in standard GR, the generalized second law (GSL) of black hole thermodynamics can be proven from the Raychaudhuri equation. Since the area law of entropy breaks down in scalar-tensor gravity, in the Jordan frame, it is worth investigating whether the Raychaudhuri equation can prove the generalized second law of thermodynamics for scalar-tensor gravity.

\subsection{Einstein frame}
Obtaining the null Raychaudhuri equation in the Einstein frame in ST theory is straightforward (the same as in standard GR). Contracting \eqref{RAMLM2TIL} with a second null vector, $l^a$, and using \eqref{THETAABTIL} one simply obtains 
\begin{align}
\t l^a\t\na_a\t\th=\t\k\t\th-\t\th_{ab}\t\th^{ab}-\t R_{ab}\t l^a\t l^b~. \label{RAYGR1}
\end{align}
Expressing $\t\th_{ab}$ in terms of the shear tensor $\t\s_{ab}$ and the expansion scalar $\t\theta$, as given by \eqref{THETAAB1TIL}, leads to equation \eqref{RAYGR1} becoming 
\begin{align}
\t l^a\t\na_a\t\th=\t\k\t\th-\t\s_{ab}\t\s^{ab}-\frac{1}{2}\t\th^2-\t R_{ab}\t l^a\t l^b~. \label{RAYGR2}
\end{align}
Next, using the equation of motion in the Einstein frame, one obtains
\begin{align}
\frac{d\t \th}{d\t\lambda}=\t \k\t \th-\t \s_{ab}\t \s^{ab}-\frac{1}{2}\t \th^2-8\pi G\t T_{ab}\t l^a\t l^b~. \label{RAYEIN}
\end{align}
To obtain the above equation we have used $\t l^a\t\na_a\t \th=d\t \th/d\t\lambda$, where $\t\lambda$ parameterizes the null geodesics, as defined by the normal $\t l^a=dx^a/d\t\lambda$~. Also note that in the above equation $\t T_{ab}$ is given as $\t T_{ab}=\t T_{ab}^{\t\phi}+\t T_{ab}^{ext}$. Now, we note that $\t T_{ab}^{\t\phi}\t l^a\t l^b=(\t l^a\p_a\t\phi)^2 \ge 0$. Therefore, if the energy-momentum tensor of the external field satisfies the null energy condition in the Einstein frame (\ie $\t T_{ab}^{ext}\t l^a\t l^b>0$), we obtain $\t T_{ab}\t l^a\t l^b>0$~.

We briefly review how the entropy increase theorem can be established in the Einstein frame from the above null Raychaudhuri equation \eqref{RAYEIN}. The process is similar to  standard GR. 
The entropy of a black hole in the Einstein frame is $\t S=\t A/4\pi$ \cite{Koga:1998un, Bhattacharya:2017pqc, Bhattacharya:2018xlq}, where $\t A$ is the surface area of the null horizon of the black hole. $\t A$ can also be written as 
\begin{align}
\t A=\int_{\mathcal{H}}\sqrt{\t q}d^2x~, \label{AREATIL}
\end{align}
where $\t q$ is the determinant of the induced metric $\t q_{ab}$~.
Now, it can be shown that 
\begin{align}
\t\th=\frac{1}{2}\t q^{ab}\tlie q_{ab}=\frac{1}{\sqrt{\t q}}\tlie\sqrt{\t q}=\frac{1}{\sqrt{\t q}}\frac{d \sqrt{\t q}}{d\t\lambda}~. \label{THETALAMTIL}
\end{align}
Using \eqref{THETALAMTIL} one can obtain the change of entropy along $\t\l^a$ (\ie $\tlie \t S=d\t S/d\t\l$) as
\begin{align}
\frac{d\t S}{d\t\lambda}=\frac{1}{4}\frac{d\t A}{d\t\lambda}=\frac{1}{4} \int_{\mathcal{H}}\sqrt{\t q}\t\th d^2x~.
\end{align}
Therefore, the entropy can only decrease (\ie $d\t S/d\t\l<0$) when $\t\th$ is negative. However, if the formation of caustics is prohibited, then from the Raychaudhuri equation \eqref{RAYEIN} one finds that $\t\th$ cannot be negative, taking in to account the null energy condition, \ie $\t T_{ab}\t l^a\t l^b>0$. From this argument one finds that the entropy must always increase.

\subsection{Jordan frame}
We now repeat the same analysis in the Jordan frame. In the Jordan frame, Bekenstein's formula for entropy breaks down since the entropy now is proportional not only to the area of the horizon, but is also proportional to the scalar field $\phi$, as suggested by Kang \cite{Kang:1996rj} (also see our previous work \cite{Bhattacharya:2017pqc, Bhattacharya:2018xlq}, where the same result was shown via different means). Thus the expression for the entropy, $S$, in the Jordan frame is
\begin{align}
S=\frac{\phi A}{4}=\frac{1}{4}\int_{\mathcal{H}}\sqrt{q}\phi d^2x~.
\end{align}
Therefore, the rate of change in entropy is
\begin{align}
\frac{dS}{d\lambda}=\frac{1}{4}\int_{\mathcal{H}}\sqrt{q}\Big(\phi\theta+l^i\na_i\phi \Big)d^2x=\frac{1}{4}\int_{\mathcal{H}}\sqrt{q}\phi\t\th d^2x~. \label{CHNGENTROPYJOR}
\end{align}
Thus, even in the Jordan frame, it is $\t\th$ and not $\th$ that determines whether the entropy should increase. This is because, as we have discussed earlier \cite{Bhattacharya:2017pqc, Bhattacharya:2018xlq}, thermodynamically the two frames are equivalent. Therefore, the appropriate Raychaudhuri equation, which is consistent with the thermodynamic description, should be defined in terms of those quantities (\ie shear, expansion scalar and energy-momentum tensor) which were obtained in the equivalent picture of the fluid-gravity dual description. From \eqref{RAYEIN}, one can obtain the expression for the Raycharudhuri equation in terms of the parameters in the Jordan frame as
\begin{align}
\frac{d\t \th}{d\lambda}=\t \k\t \th-\frac{1}{\phi^2} \Sigma_{ab} \Sigma^{ab}-\frac{1}{2}\t \th^2-8\pi G\t T_{ab}l^al^b~, \label{RAYJOR}
\end{align}
where $\t T_{ab}=\t T_{ab}^{\t\phi}+(T_{ab}^{ext}/\phi)$~. To obtain the above equation, we have used $d\t\theta/d\t\lambda=\t l^a\p_a\t\theta= l^a\p_a\t\theta=d\t\theta/d\lambda$~. We again note that $\t T_{ab}^{\t\phi}l^al^b=(l^a\p_a\t\phi)^2 \ge 0$. Therefore, when the external matter field satisfies the null energy condition (\ie $T_{ab}^{ext}l^al^b>0$), we obtain $\t T_{ab}l^al^b>0$~. Then following the arguments as in the Einstein frame, one can prove that $\t\th$ is always positive, which proves the entropy increase theorem in the Jordan frame.

In this section we have found that it is $\t\th$ which comes into play in obtaining the entropy increase theorem in the Jordan frame. This is consistent with the fact that the two frames are thermodynamically equivalent. In addition, we have obtained the null Raychaudhuri equation in each frame which is consistent with the thermodynamic description in scalar-tensor theory.



\section{Tidal force equation}
The Raychaudhuri equation discussed in the previous section describes how the trace part of $\th_{ab}$ or $\t\th_{ab}$ (\ie $\th$ or $\t\th$) evolves in the spacetime. On the other hand, the tidal force equation describes the evolution of the traceless part of $\th_{ab}$ or $\t\th_{ab}$ (\ie $\sigma_{ab}$ or $\t\sigma_{ab}$) \cite{Price:1986yy}. In this section, we shall discuss the expression of the tidal force in the two frames. 
In obtaining the tidal force equation, the equation of motion is not used. Therefore, the standard expression for the tidal force equation from GR is also valid for scalar tensor theory. 

In our equivalent fluid-gravity dual description, it is $\Sigma_{ab}$ (and not $\sigma_{ab}$) which we have identified as the shear tensor in the Jordan frame. $\Sigma_{ab}$ is also conformally invariant, since $\Sigma_{ab}=\t\sigma_{ab}$. Thus, in the Jordan frame, we have to examine how $\Sigma_{ab}$ evolves in the spacetime. The standard expression of the tidal force equation \cite{Price:1986yy, Gourgoulhon:2005ng} valid in the Einstein frame, is given by
\begin{align}
\t q^i_a\t q^j_b(\tlie\t\s_{ij})-\t\k\t\s_{ab}-\t q_{ab}\t\s^{ij}\t\s_{ij}=-\t q^i_a\t q^j_b\t C_{minj}\t l^m\t l^n~, \label{TIDALEIN}
\end{align}
where $\t C_{abcd}$ is the Weyl tensor -- the traceless part of the Riemann tensor $\t R_{abcd}$~. The explicit form of $\t C_{abcd}$ is 
\begin{align}
\t C_{abcd}=\t R_{abcd}-\frac{2}{n-2}\Big(\t g_{a[c}\t R_{d]b}-\t g_{b[c}\t R_{d]a}\Big)
\no 
\\
+\frac{2}{(n-1)(n-2)}\Big(\t g_{a[c}\t g_{d]b}\Big)~, \label{WEYL}
\end{align}
where $n$ is the dimension of the whole manifold.
The RHS of \eqref{TIDALEIN} gives the measure of the tidal force on the two-surface $\S_t$ . 

We now obtain the the tidal force equation in the Jordan frame. Using the conformal transformation \eqref{GAB} and \eqref{PHI} we can obtain, from \eqref{TIDALEIN},
\begin{align}
 q^i_a q^j_b(\lie\Sigma_{ij})-\t\k\Sigma_{ab}-\phi q_{ab}\Sigma^{ij}\Sigma_{ij}=-\phi q^i_a q^j_b C_{minj} l^m l^n~, \label{TIDALJOR}
\end{align}
The above form of the tidal force equation is consistent with the equivalent fluid-gravity picture.

However, the usual the tidal force equation of GR is consistent with the inequivalent fluid-gravity picture. In this case, the expression of tidal force equation is
\begin{align}
q^i_a q^j_b(\lie\s_{ij})-\k\s_{ab}- q_{ab}\s^{ij}\s_{ij}=- q^i_aq^j_b C_{minj}l^m l^n~. \label{TIDALJORINEQ}
\end{align}

 
 \section{Summary and Conclusions} 
 The connection of the field equation of the gravity with fluid dynamics (\ie that the Einstein equation can be identified with the Navier-Stokes equation  by projecting the GR equation onto a null surface) is a remarkable discovery due to Damour \cite{DAMOUR} . Later, this fluid-gravity connection was established in several ways \cite{Price:1986yy, Gourgoulhon:2005ng, Padmanabhan:2010rp, Kolekar:2011gw, Parikh:1997ma, Bredberg:2011jq, Bhattacharyya:2008kq, Huang:2011he, Chirco:2011ex, Bai:2012ci, Bredberg:2010ky, Cai:2012mg, Zou:2013ix, Hu:2013lua, Cai:2011xv, Huang:2011kj, Anninos:2011zn, Ling:2013kua, Eling:2012ni, Berkeley:2012kz, Lysov:2017cmc, Wu:2013aov, Compere:2011dx, De:2019wok, De:2018zxo, Chatterjee:2010gp} and this connection is now considered as a remarkable feature of general relativity. Over the last few decades, several alternative theories of gravity have developed as extensions/substitutions of GR. The question, which motivated the analysis in this paper, was whether the fluid-gravity connection is a special characteristic of standard GR only, or can one obtain a similar connection for these alternative gravity theories. Also it is of interest to determine how the expressions of the fluid parameters in these alternative gravity theories look, and to what extent they differ from standard GR. In this work, we have obtained the fluid-gravity analogy in the scalar-tensor theory of gravity, which is one of the most promising theories among various alternative theories of gravity. Moreover, since any $f(R)$ gravity theory can be written in terms of a corresponding scalar-tensor theory, this work implies that the present analysis is relevant for establishing the fluid-gravity correspondence in higher curvature gravity theories. In addition, scalar-tensor theory is described in both the Jordan frame and the Einstein frame. These two frames are conformally connected to each other. The physical equivalence of these two conformally connected frames has been a matter of debate for several decades. From the thermodynamic viewpoint, the two frames are physically equivalent. However, from the fluid-gravity viewpoint the above analysis shows that no definite conclusion can be drawn using the fluid/gravity analysis. Further investigation is required to figure out whether the fluid parameters are equivalent in the two frames.
 
In this paper we first provided a brief overview of the scalar-tensor theory of gravity in both the Jordan and Einstein frames. Second, we briefly described the geometry of the null-surface in the context of ($1+3$) spacetime foliation, an idea which is necessary in order to obtain the projected form of the gravitational field equation onto the null surface. Then, we obtained the DNS equation in the Einstein frame, following a similar approach to that used in standard GR. We found that the complete form of the DNS equation could be obtained even without considering the external matter field, unlike the case in standard GR. Later we showed that one cannot obtain the DNS equation in the Jordan frame by following the same procedure as in standard GR. In particular, in the Jordan frame, one finds that there are two different cases to get the DNS equation. In the first case, we have shown that one can obtain the DNS equation along with an additional Coriolis-like force term. This picture is consistent with some recent work in the framework of $f(R)$ gravity. Moreover, in this first case, the ratio $\eta/s$ saturates the KSS bound. However, in this case, the fluid parameters are not equivalent to those of the Einstein frame. Later, we showed that an equivalent case is also possible in the Jordan frame. In this second case the DNS equation can be obtained in terms of fluid variables, which are equivalent to those of the Einstein frame. In this method, the Coriolis-like force term does not appear in the DNS equation. However, as we have mentioned earlier, the equivalent picture may violate the KSS bound when $\phi>1$. Thus the violation of the KSS bound might be allowed for scalar-tensor theory. However, one has to investigate whether the theory remains causally well-behaved for those values of $\phi$~. In addition, we believe, more investigation is required in this regard to find out which of the two pictures is more consistent. The present fluid-gravity analysis does not rule out either of the two pictures.
 
  After obtaining the DNS equation in the Einstein and Jordan frames, we discussed the Raychaudhuri equation in the two frames in the context of the generalized second law. As shown in this work, the second law of thermodynamics can be obtained from the Raychaudhuri equation. However, in the Jordan frame, the evolution of the expansion scalar $\theta$ does not determine the entropy increase theorem, but rather it is determined by $\t\theta$. Therefore, we identify the proper Raychaudhuri equation as the evolution of $\t\th$ in the Jordan frame. Finally, we showed how the shear tensor evolves in the spacetime by obtaining the tidal force equation in the two frames. We mentioned how the tidal force equation of  standard GR continues to be valid for the Einstein frame of ST gravity. The same expression of the tidal equation is also consistent with our inequivalent picture of the fluid-gravity duality in the Jordan frame. Additionally, we also obtained the tidal force equation which is consistent with the second (equivalent) case of the fluid-dynamic description in the Jordan frame.
 
The analysis in this paper indicates that the fluid-gravity correspondence is a general characteristic of gravity theories, similar to the correspondence between thermodynamics and gravity. This connection with fluid dynamics is not limited to standard GR. In the present case, we have shown that this correspondence with fluid dynamics also works with scalar-tensor gravity. This connection between fluid dynamics and scalar-tensor theory, also implies that this correspondence exists for $f(R)$ gravity, since $f(R)$ gravity can be described as a scalar-tensor gravity. Our analysis also suggests that, unlike the thermodynamic case, where the two frames are exactly equivalent, from the fluid-gravity perspective it is possible to obtain both equivalent as well as inequivalent cases. In our analysis, we have provided arguments for/against each of the pictures (case 1 and case 2) and have pointed out that both pictures have some arguments in favor of their validity. Thus, the present analysis should be useful in the context of the fluid-gravity analogy and also in the theory of scalar-tensor gravity.
 
 \vskip 4mm
{\section*{Acknowledgments}}
\noindent
One of the authors (KB) sincerely thanks to Shounak De and Sumit Dey for some useful discussions. The research of KB is supported by Institute of International Education, US by the Fulbright-Nehru Doctoral Fellowship (Grantee ID: E0609748)~.
 \vskip 4mm
  
 


\begin{thebibliography}{99}

\bibitem{Riess:2001gk} 
  A.~G.~Riess {\it et al.} [Supernova Search Team Collaboration],
  ``The farthest known supernova: support for an accelerating universe and a glimpse of the epoch of deceleration,''
  Astrophys.\ J.\  {\bf 560}, 49 (2001)
  [astro-ph/0104455].
 
\bibitem{Riess:2004nr} 
  A.~G.~Riess {\it et al.} [Supernova Search Team Collaboration],
  ``Type Ia supernova discoveries at z $>$ 1 from the Hubble Space Telescope: Evidence for past deceleration and constraints on dark energy evolution,''
  Astrophys.\ J.\  {\bf 607}, 665 (2004)
  [astro-ph/0402512].
 
 \bibitem{KNOP}
 R.~A.~Knop et al.,
 ``New Constraints on $\Omega_M$, $\Omega_{\Lambda}$, and $w$ from an Independent Set of 11 High-Redshift Supernovae Observed with the Hubble Space Telescope'',
  The Astrophysical Journal, 2003, {\bf 598}, 102-137.
 
\bibitem{Perlmutter:1998np} 
  S.~Perlmutter {\it et al.} [Supernova Cosmology Project Collaboration],
  ``Measurements of Omega and Lambda from 42 high redshift supernovae,''
  Astrophys.\ J.\  {\bf 517}, 565 (1999),
  [astro-ph/9812133].
  
  
\bibitem{Tonry:2003zg} 
  J.~L.~Tonry {\it et al.} [Supernova Search Team Collaboration],
  ``Cosmological results from high-z supernovae,''
  Astrophys.\ J.\  {\bf 594}, 1 (2003),
  [astro-ph/0305008].

\bibitem{Barris:2003dq} 
  B.~J.~Barris {\it et al.},
  ``23 High redshift supernovae from the IFA Deep Survey: Doubling the SN sample at $z > 0.7$,''
  Astrophys.\ J.\  {\bf 602}, 571 (2004),
  [astro-ph/0310843].
  
  
\bibitem{Perlmutter:1997zf} 
  S.~Perlmutter {\it et al.} [Supernova Cosmology Project Collaboration],
  ``Discovery of a supernova explosion at half the age of the Universe and its cosmological implications,''
  Nature {\bf 391}, 51 (1998)
  [astro-ph/9712212].

\bibitem{Riess:1998cb} 
  A.~G.~Riess {\it et al.} [Supernova Search Team Collaboration],
  ``Observational evidence from supernovae for an accelerating universe and a cosmological constant,''
  Astron.\ J.\  {\bf 116}, 1009 (1998)
  [astro-ph/9805201].

\bibitem{RIESS4}
Adam~G.~Riess et al.,
``Is there an Indication of Evolution of Type Ia Supernovaefrom their Rise Times?",
The Astronomical Journal, 1999, {\bf 118}, 2668-2674.

\bibitem{Starobinsky:1979ty} 
  A.~A.~Starobinsky,
  ``Spectrum of relict gravitational radiation and the early state of the universe,''
  JETP Lett.\  {\bf 30}, 682 (1979)
  [Pisma Zh.\ Eksp.\ Teor.\ Fiz.\  {\bf 30}, 719 (1979)]. 
 
\bibitem{Guth:1980zm} 
  A.~H.~Guth,
  ``The Inflationary Universe: A Possible Solution to the Horizon and Flatness Problems,''
  Phys.\ Rev.\ D {\bf 23}, 347 (1981)
  [Adv.\ Ser.\ Astrophys.\ Cosmol.\  {\bf 3}, 139 (1987)].
  
\bibitem{Linde:1981mu} 
  A.~D.~Linde,
  ``A New Inflationary Universe Scenario: A Possible Solution of the Horizon, Flatness, Homogeneity, Isotropy and Primordial Monopole Problems,''
  Phys.\ Lett.\  {\bf 108B}, 389 (1982)
  [Adv.\ Ser.\ Astrophys.\ Cosmol.\  {\bf 3}, 149 (1987)].
  
\bibitem{Damour:2002mi} 
  T.~Damour, F.~Piazza and G.~Veneziano,
  ``Runaway dilaton and equivalence principle violations,''
  Phys.\ Rev.\ Lett.\  {\bf 89}, 081601 (2002)
  [gr-qc/0204094].

\bibitem{Clifton:2011jh} 
  T.~Clifton, P.~G.~Ferreira, A.~Padilla and C.~Skordis,
  ``Modified Gravity and Cosmology,''
  Phys.\ Rept.\  {\bf 513}, 1 (2012)
  [arXiv:1106.2476 [astro-ph.CO]].
  
 \bibitem{Nojiri:2017ncd} 
  S.~Nojiri, S.~D.~Odintsov and V.~K.~Oikonomou,
  ``Modified Gravity Theories on a Nutshell: Inflation, Bounce and Late-time Evolution,''
  Phys.\ Rept.\  {\bf 692}, 1 (2017)
  [arXiv:1705.11098 [gr-qc]].
  
\bibitem{Callan:1985ia} 
  C.~G.~Callan, Jr., E.~J.~Martinec, M.~J.~Perry and D.~Friedan,
  ``Strings in Background Fields,''
  Nucl.\ Phys.\ B {\bf 262}, 593 (1985).
  
\bibitem{EspositoFarese:2003ze} 
  G.~Esposito-Farese,
  ``Scalar tensor theories and cosmology and tests of a quintessence Gauss-Bonnet coupling,''
  gr-qc/0306018.
  
\bibitem{Elizalde:2004mq} 
  E.~Elizalde, S.~Nojiri and S.~D.~Odintsov,
  ``Late-time cosmology in (phantom) scalar-tensor theory: Dark energy and the cosmic speed-up,''
  Phys.\ Rev.\ D {\bf 70}, 043539 (2004)
  [hep-th/0405034].
  
\bibitem{Saridakis:2016ahq}
  E.~N.~Saridakis and M.~Tsoukalas,
  ``Cosmology in new gravitational scalar-tensor theories,''
  Phys.\ Rev.\ D {\bf 93} (2016) no.12,  124032
  [arXiv:1601.06734 [gr-qc]].
  
\bibitem{Crisostomi:2016czh} 
  M.~Crisostomi, K.~Koyama and G.~Tasinato,
  ``Extended Scalar-Tensor Theories of Gravity,''
  JCAP {\bf 1604}, no. 04, 044 (2016)
  [arXiv:1602.03119 [hep-th]].
  
\bibitem{Langlois:2017dyl} 
  D.~Langlois, R.~Saito, D.~Yamauchi and K.~Noui,
  ``Scalar-tensor theories and modified gravity in the wake of GW170817,''
  arXiv:1711.07403 [gr-qc].
  
  
  
\bibitem{Bhattacharya:2017pqc} 
  K.~Bhattacharya and B.~R.~Majhi,
  ``Fresh look at the scalar-tensor theory of gravity in Jordan and Einstein frames from undiscussed standpoints,''
  Phys.\ Rev.\ D {\bf 95}, no. 6, 064026 (2017)
  [arXiv:1702.07166 [gr-qc]].
  
\bibitem{Bhattacharya:2018xlq} 
  K.~Bhattacharya, A.~Das and B.~R.~Majhi,
  ``Noether and Abbott-Deser-Tekin conserved quantities in scalar-tensor theory of gravity both in Jordan and Einstein frames,''
  Phys.\ Rev.\ D {\bf 97}, no. 12, 124013 (2018)
  [arXiv:1803.03771 [gr-qc]].
  
  \bibitem{DAMOUR}
  T.~Damour (1979),
 {\it ``Quelques propri\'et\'es m\'ecaniques, \'electromagn\'etiques, thermodynamiques et
quantiques des trous noirs''}, Th\`ese de doctorat d\'Etat, Universit\'e Paris 6;
T.~Damour (1982), ``Surface effects in black hole physics, Proceedings of the Second Marcel Grossmann Meeting on General Relativity'', Ed. R. Ruffini, North Holland, p. 587.

  
\bibitem{Price:1986yy} 
  R.~H.~Price and K.~S.~Thorne,
  ``Membrane Viewpoint on Black Holes: Properties and Evolution of the Stretched Horizon,''
  Phys.\ Rev.\ D {\bf 33}, 915 (1986).
  
\bibitem{Gourgoulhon:2005ng} 
  E.~Gourgoulhon and J.~L.~Jaramillo,
  ``A 3+1 perspective on null hypersurfaces and isolated horizons,''
  Phys.\ Rept.\  {\bf 423}, 159 (2006)
  [gr-qc/0503113].
  
  
\bibitem{Padmanabhan:2010rp} 
  T.~Padmanabhan,
  ``Entropy density of spacetime and the Navier-Stokes fluid dynamics of null surfaces,''
  Phys.\ Rev.\ D {\bf 83}, 044048 (2011)
  [arXiv:1012.0119 [gr-qc]].
  
\bibitem{Kolekar:2011gw} 
  S.~Kolekar and T.~Padmanabhan,
  ``Action Principle for the Fluid-Gravity Correspondence and Emergent Gravity,''
  Phys.\ Rev.\ D {\bf 85}, 024004 (2012)
  [arXiv:1109.5353 [gr-qc]].
  
  
  
  
  
\bibitem{Parikh:1997ma} 
  M.~Parikh and F.~Wilczek,
  ``An Action for black hole membranes,''
  Phys.\ Rev.\ D {\bf 58}, 064011 (1998)
  [gr-qc/9712077].
  

  
\bibitem{Bredberg:2011jq} 
  I.~Bredberg, C.~Keeler, V.~Lysov and A.~Strominger,
  ``From Navier-Stokes To Einstein,''
  JHEP {\bf 1207}, 146 (2012)
  [arXiv:1101.2451 [hep-th]].
  
\bibitem{Bhattacharyya:2008kq} 
  S.~Bhattacharyya, S.~Minwalla and S.~R.~Wadia,
  ``The Incompressible Non-Relativistic Navier-Stokes Equation from Gravity,''
  JHEP {\bf 0908}, 059 (2009)
  [arXiv:0810.1545 [hep-th]].
  
\bibitem{Huang:2011he} 
  T.~Z.~Huang, Y.~Ling, W.~J.~Pan, Y.~Tian and X.~N.~Wu,
  ``From Petrov-Einstein to Navier-Stokes in Spatially Curved Spacetime,''
  JHEP {\bf 1110}, 079 (2011)
  [arXiv:1107.1464 [gr-qc]].
  
\bibitem{Chirco:2011ex} 
  G.~Chirco, C.~Eling and S.~Liberati,
  ``Higher Curvature Gravity and the Holographic fluid dual to flat spacetime,''
  JHEP {\bf 1108}, 009 (2011)
  [arXiv:1105.4482 [hep-th]].
  
\bibitem{Bai:2012ci} 
  X.~Bai, Y.~P.~Hu, B.~H.~Lee and Y.~L.~Zhang,
  ``Holographic Charged Fluid with Anomalous Current at Finite Cutoff Surface in Einstein-Maxwell Gravity,''
  JHEP {\bf 1211}, 054 (2012)
  [arXiv:1207.5309 [hep-th]].
  
\bibitem{Bredberg:2010ky} 
  I.~Bredberg, C.~Keeler, V.~Lysov and A.~Strominger,
  ``Wilsonian Approach to Fluid/Gravity Duality,''
  JHEP {\bf 1103}, 141 (2011)
  [arXiv:1006.1902 [hep-th]].
  
\bibitem{Chatterjee:2010gp} 
  S.~Chatterjee, M.~Parikh and S.~Sarkar,
  ``The Black Hole Membrane Paradigm in f(R) Gravity,''
  Class.\ Quant.\ Grav.\  {\bf 29}, 035014 (2012)
  [arXiv:1012.6040 [hep-th]].
  
\bibitem{Cai:2012mg} 
  R.~G.~Cai, T.~J.~Li, Y.~H.~Qi and Y.~L.~Zhang,
  ``Incompressible Navier-Stokes Equations from Einstein Gravity with Chern-Simons Term,''
  Phys.\ Rev.\ D {\bf 86}, 086008 (2012)
  [arXiv:1208.0658 [hep-th]].
  
\bibitem{Zou:2013ix} 
  D.~C.~Zou, S.~J.~Zhang and B.~Wang,
  ``Holographic charged fluid dual to third order Lovelock gravity,''
  Phys.\ Rev.\ D {\bf 87}, no. 8, 084032 (2013)
  [arXiv:1302.0904 [hep-th]].
  
\bibitem{Hu:2013lua} 
  Y.~P.~Hu, Y.~Tian and X.~N.~Wu,
  ``Bulk Viscosity of dual Fluid at Finite Cutoff Surface via Gravity/Fluid correspondence in Einstein-Maxwell Gravity,''
  Phys.\ Lett.\ B {\bf 732}, 298 (2014)
  [arXiv:1311.3891 [hep-th]].
  
\bibitem{Cai:2011xv} 
  R.~G.~Cai, L.~Li and Y.~L.~Zhang,
  ``Non-Relativistic Fluid Dual to Asymptotically AdS Gravity at Finite Cutoff Surface,''
  JHEP {\bf 1107}, 027 (2011)
  [arXiv:1104.3281 [hep-th]].
  
\bibitem{Huang:2011kj} 
  T.~Z.~Huang, Y.~Ling, W.~J.~Pan, Y.~Tian and X.~N.~Wu,
  ``Fluid/gravity duality with Petrov-like boundary condition in a spacetime with a cosmological constant,''
  Phys.\ Rev.\ D {\bf 85}, 123531 (2012)
  [arXiv:1111.1576 [hep-th]].
  
\bibitem{Anninos:2011zn} 
  D.~Anninos, T.~Anous, I.~Bredberg and G.~S.~Ng,
  ``Incompressible Fluids of the de Sitter Horizon and Beyond,''
  JHEP {\bf 1205}, 107 (2012)
  [arXiv:1110.3792 [hep-th]].
  
\bibitem{Ling:2013kua} 
  Y.~Ling, C.~Niu, Y.~Tian, X.~N.~Wu and W.~Zhang,
  ``Note on the Petrov-like boundary condition at finite cutoff surface in gravity/fluid duality,''
  Phys.\ Rev.\ D {\bf 90}, no. 4, 043525 (2014)
  [arXiv:1306.5633 [gr-qc]].
  
\bibitem{Eling:2012ni} 
  C.~Eling, A.~Meyer and Y.~Oz,
  ``The Relativistic Rindler Hydrodynamics,''
  JHEP {\bf 1205}, 116 (2012)
  [arXiv:1201.2705 [hep-th]].
  
\bibitem{Berkeley:2012kz} 
  J.~Berkeley and D.~S.~Berman,
  ``The Navier-Stokes equation and solution generating symmetries from holography,''
  JHEP {\bf 1304}, 092 (2013)
  [arXiv:1211.1983 [hep-th]].
  
\bibitem{Lysov:2017cmc} 
  V.~Lysov,
  ``Dual Fluid for the Kerr Black Hole,''
  JHEP {\bf 1806}, 080 (2018)
  [arXiv:1712.08079 [hep-th]].
  
\bibitem{Wu:2013aov} 
  X.~Wu, Y.~Ling, Y.~Tian and C.~Zhang,
  ``Fluid/Gravity Correspondence For General Non-rotating Black Holes,''
  Class.\ Quant.\ Grav.\  {\bf 30}, 145012 (2013)
  [arXiv:1303.3736 [hep-th]].
  
\bibitem{Compere:2011dx} 
  G.~Compere, P.~McFadden, K.~Skenderis and M.~Taylor,
  ``The Holographic fluid dual to vacuum Einstein gravity,''
  JHEP {\bf 1107}, 050 (2011)
  [arXiv:1103.3022 [hep-th]].
  
\bibitem{De:2019wok} 
  S.~De, S.~Dey and B.~R.~Majhi,
  ``Effective metric in fluid-gravity duality through parallel transport: a proposal,''
  Phys.\ Rev.\ D {\bf 99}, no. 12, 124024 (2019)
  [arXiv:1901.05735 [hep-th]].
  
\bibitem{De:2018zxo} 
  S.~De and B.~R.~Majhi,
  ``Fluid description of gravity on a timelike cut-off surface: beyond Navier-Stokes equation,''
  JHEP {\bf 1901}, 044 (2019)
  [arXiv:1810.07017 [hep-th]].
  
\bibitem{Faraoni:1998qx} 
  V.~Faraoni, E.~Gunzig and P.~Nardone,
  ``Conformal transformations in classical gravitational theories and in cosmology,''
  Fund.\ Cosmic Phys.\  {\bf 20}, 121 (1999)
  [gr-qc/9811047].
  
 \bibitem{Quiros:2018ryt} 
  I.~Quiros and R.~De Arcia,
  ``On local scale invariance and the questionable theoretical basis of the conformal transformations' issue,''
  arXiv:1811.02458 [gr-qc].
  
\bibitem{Kamenshchik:2014waa} 
  A.~Y.~Kamenshchik and C.~F.~Steinwachs,
  ``Question of quantum equivalence between Jordan frame and Einstein frame,''
  Phys.\ Rev.\ D {\bf 91}, no. 8, 084033 (2015)
  [arXiv:1408.5769 [gr-qc]].
 
  
\bibitem{Banerjee:2016lco} 
  N.~Banerjee and B.~Majumder,
  ``A question mark on the equivalence of Einstein and Jordan frames,''
  Phys.\ Lett.\ B {\bf 754}, 129 (2016)
  [arXiv:1601.06152 [gr-qc]].
  
 \bibitem{Ruf:2017xon} 
  M.~S.~Ruf and C.~F.~Steinwachs,
  ``Quantum equivalence of $f(R)$ gravity and scalar-tensor theories,''
  Phys.\ Rev.\ D {\bf 97}, no. 4, 044050 (2018)
  [arXiv:1711.07486 [gr-qc]].
  
\bibitem{Frion:2018oij} 
  E.~Frion and C.~R.~Almeida,
  ``Affine quantization of the Brans-Dicke theory: Smooth bouncing and the equivalence between the Einstein and Jordan frames,''
  Phys.\ Rev.\ D {\bf 99}, no. 2, 023524 (2019)
  [arXiv:1810.00707 [gr-qc]].
  
\bibitem{Karam:2017zno} 
  A.~Karam, T.~Pappas and K.~Tamvakis,
  ``Frame-dependence of higher-order inflationary observables in scalar-tensor theories,''
  Phys.\ Rev.\ D {\bf 96}, no. 6, 064036 (2017)
  [arXiv:1707.00984 [gr-qc]].
    
\bibitem{Bahamonde:2017kbs} 
  S.~Bahamonde, S.~D.~Odintsov, V.~K.~Oikonomou and P.~V.~Tretyakov,
  ``Deceleration versus acceleration universe in different frames of $F(R)$ gravity,''
  Phys.\ Lett.\ B {\bf 766}, 225 (2017)
  [arXiv:1701.02381 [gr-qc]].
  
\bibitem{Karam:2018squ} 
  A.~Karam, A.~Lykkas and K.~Tamvakis,
  ``Frame-invariant approach to higher-dimensional scalar-tensor gravity,''
  Phys.\ Rev.\ D {\bf 97}, no. 12, 124036 (2018)
  [arXiv:1803.04960 [gr-qc]].
  
  
  

  

  
  
\bibitem{Brustein:2008cg} 
  R.~Brustein and A.~J.~M.~Medved,
  ``The Ratio of shear viscosity to entropy density in generalized theories of gravity,''
  Phys.\ Rev.\ D {\bf 79}, 021901 (2009)
  [arXiv:0808.3498 [hep-th]].

  
\bibitem{Kovtun:2004de} 
  P.~Kovtun, D.~T.~Son and A.~O.~Starinets,
  ``Viscosity in strongly interacting quantum field theories from black hole physics,''
  Phys.\ Rev.\ Lett.\  {\bf 94}, 111601 (2005)
  [hep-th/0405231].
  
\bibitem{Faraoni:1999hp} 
  V.~Faraoni and E.~Gunzig,
  ``Einstein frame or Jordan frame?,''
  Int.\ J.\ Theor.\ Phys.\  {\bf 38}, 217 (1999)
  [astro-ph/9910176].
  
\bibitem{Koga:1998un} 
  J.~i.~Koga and K.~i.~Maeda,
  ``Equivalence of black hole thermodynamics between a generalized theory of gravity and the Einstein theory,''
  Phys.\ Rev.\ D {\bf 58}, 064020 (1998)
  [gr-qc/9803086].
  
\bibitem{Faraoni:2010yi} 
  V.~Faraoni,
  ``Black hole entropy in scalar-tensor and f(R) gravity: An Overview,''
  Entropy {\bf 12}, 1246 (2010)
  [arXiv:1005.2327 [gr-qc]].
 
\bibitem{Brigante:2007nu}
M.~Brigante, H.~Liu, R.~C.~Myers, S.~Shenker and S.~Yaida,
``Viscosity Bound Violation in Higher Derivative Gravity,''
Phys. Rev. D \textbf{77}, 126006 (2008)
[arXiv:0712.0805 [hep-th]].

\bibitem{Brigante:2008gz}
M.~Brigante, H.~Liu, R.~C.~Myers, S.~Shenker and S.~Yaida,
``The Viscosity Bound and Causality Violation,''
Phys. Rev. Lett. \textbf{100}, 191601 (2008)
[arXiv:0802.3318 [hep-th]].

\bibitem{Kats:2007mq}
Y.~Kats and P.~Petrov,
``Effect of curvature squared corrections in AdS on the viscosity of the dual gauge theory,''
JHEP \textbf{01}, 044 (2009)
[arXiv:0712.0743 [hep-th]].

\bibitem{Son:2007vk}
D.~T.~Son and A.~O.~Starinets,
``Viscosity, Black Holes, and Quantum Field Theory,''
Ann. Rev. Nucl. Part. Sci. \textbf{57}, 95-118 (2007)
[arXiv:0704.0240 [hep-th]].

\bibitem{Buchel:2004di}
A.~Buchel, J.~T.~Liu and A.~O.~Starinets,
``Coupling constant dependence of the shear viscosity in N=4 supersymmetric Yang-Mills theory,''
Nucl. Phys. B \textbf{707}, 56-68 (2005)
[arXiv:hep-th/0406264 [hep-th]].

\bibitem{Camanho:2014apa}
X.~O.~Camanho, J.~D.~Edelstein, J.~Maldacena and A.~Zhiboedov,
``Causality Constraints on Corrections to the Graviton Three-Point Coupling,''
JHEP \textbf{02}, 020 (2016)
[arXiv:1407.5597 [hep-th]].

\bibitem{Papallo:2015rna}
G.~Papallo and H.~S.~Reall,
``Graviton time delay and a speed limit for small black holes in Einstein-Gauss-Bonnet theory,''
JHEP \textbf{11}, 109 (2015)
[arXiv:1508.05303 [gr-qc]].

\bibitem{Kang:1996rj} 
  G.~Kang,
  ``On black hole area in Brans-Dicke theory,''
  Phys.\ Rev.\ D {\bf 54}, 7483 (1996)
  [gr-qc/9606020].
  
    
\end{thebibliography}
\end{document}